\documentclass[acmtog]{acmart}

\usepackage{enumitem}
\usepackage{colortbl}

\AtBeginDocument{%
  }

\setcopyright{acmlicensed}
\copyrightyear{2026}
\acmYear{2026}
\setcopyright{cc}
\setcctype{by}
\acmConference[SA Conference Papers '26]{SIGGRAPH Asia 2026 Conference Papers}{December 01--04, 2026}{Kuala Lumpur, Malaysia}
\acmBooktitle{SIGGRAPH Asia 2026 Conference Papers (SA Conference Papers '26), December 01--04, 2026, Kuala Lumpur, Malaysia}
\acmDOI{10.1145/3829340.3842366}
\acmISBN{979-8-4007-2842-6/2026/12}

\begin{document}

\title{TalkFlow: Structured Latent Flow Matching for Speech Driven Holistic Motion Generation}

\author{Xinmu Wang}
\email{xinmuwang@cs.stonybrook.edu}
\orcid{0000-0001-5655-4342}
\affiliation{%
  \institution{Stony Brook University}
  \city{Stony Brook}
  \state{NY}
  \country{USA}
}

\author{Fumi Honda}
\email{fhonda@cs.stonybrook.edu}
\orcid{0009-0006-3855-4094}
\affiliation{%
  \institution{Stony Brook University}
  \city{Stony Brook}
  \state{NY}
  \country{USA}
}

\author{Xiang Gao}
\email{gao2@cs.stonybrook.edu}
\orcid{0009-0005-4385-3630}
\affiliation{%
  \institution{Stony Brook University}
  \city{Stony Brook}
  \state{NY}
  \country{USA}
}

\author{Xianfeng Gu}
\email{gu@cs.stonybrook.edu}
\orcid{0000-0001-8226-5851}
\affiliation{%
  \institution{Stony Brook University}
  \city{Stony Brook}
  \state{NY}
  \country{USA}
}

\author{Xiyun Song}
\email{xsong@futurewei.com}
\orcid{0009-0006-3114-7322}
\affiliation{%
  \institution{Futurewei Technologies Inc}
  \city{Santa Clara}
  \state{CA}
  \country{USA}
}

\author{Heather Yu}
\email{hyu@futurewei.com}
\orcid{0009-0008-4793-7585}
\affiliation{%
  \institution{Futurewei Technologies Inc}
  \city{Santa Clara}
  \state{CA}
  \country{USA}
}

\renewcommand{\shortauthors}{Wang et al.}

\begin{abstract}
Holistic co-speech motion generation aims to synthesize coordinated facial expressions, hand gestures, and body movements from speech, which is challenging due to heterogeneous part-specific dynamics, long-range temporal coherence, and cross-part coordination. Directly modeling high-dimensional motion can be unstable, while overly compact or unstructured latent spaces may lose fine-grained motion details. We present \textbf{TalkFlow}, a structured latent flow framework for speech-driven holistic motion generation. TalkFlow learns a continuous three-stream latent representation for upper-body motion, lower-body motion, and facial expressions, preserving part-specific dynamics without forcing heterogeneous motion components into a single latent space. To maintain holistic coordination, lightweight cross-stream interactions are incorporated in both latent representation learning and flow-based generation. Conditioned on multi-level speech features, TalkFlow performs speech-conditioned latent flow matching, learning a continuous transport process from source latents to realistic motion latents. Experiments on the BEAT2 dataset demonstrate strong performance across facial expression accuracy, body gesture quality, and speech-motion synchronization. Qualitative comparisons and user studies further show that TalkFlow generates more natural, expressive, and speech-aligned holistic motion than state-of-the-art methods.
The project page is available at \url{https://xinmuwang.github.io/TalkFlow/}.

\end{abstract}



\begin{teaserfigure}
  \centering
  \includegraphics[width=1\textwidth]{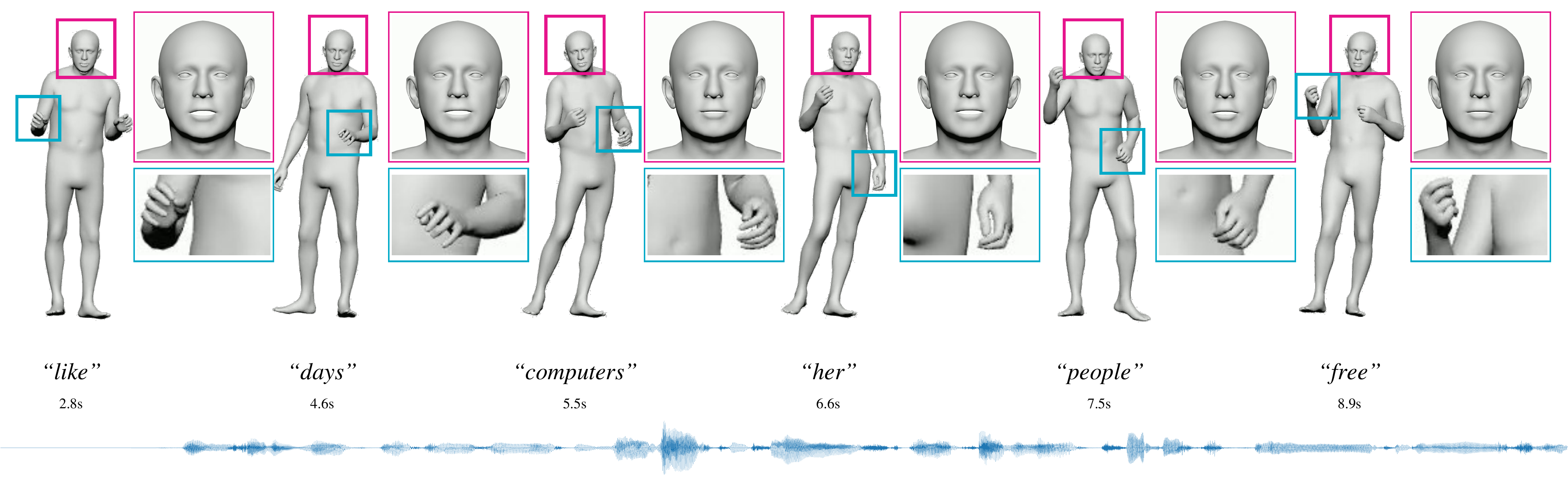}
    \caption{
    \textbf{TalkFlow} generates coordinated body, hand, and facial motion from speech via structured latent flow matching. 
    Conditioned on speech audio and aligned text features, our method synthesizes speech-aligned holistic motion: each frame shows the generated body pose with zoomed views of the face and hand. 
    Frames are sampled at the onset of content words across a $9$-second clip, with the corresponding waveform shown at the bottom.
    }
  \label{fig:test_teaser}
  \Description{Test teaser description.}
\end{teaserfigure}

\maketitle

\section{Introduction}
\label{sec:intro}

Generating expressive human motion from speech is a central problem for digital avatars, virtual agents, augmented and virtual reality, and embodied communication. With parametric human models such as SMPL-X~\cite{pavlakos2019expressive}, holistic human motion can be represented in a unified form that includes body pose, hand articulation, and facial expressions. Recent co-speech datasets and benchmarks, such as BEAT2~\cite{liu2024emage}, further make it possible to study speech-driven 3D human motion at scale. However, generating realistic holistic motion remains challenging because speech-driven facial expressions, hand gestures, upper-body movements, and lower-body posture exhibit heterogeneous dynamics while requiring temporal coherence and cross-part coordination.

Existing co-speech motion generation methods have made significant progress. Early learning-based approaches mainly relied on deterministic regression~\cite{habibie2021learning, ng2022learning}, which often averages over multiple plausible gestures. More recent methods improve diversity and realism through discrete latent modeling, masked modeling, diffusion, or sequence modeling architectures~\cite{yi2023generating, liu2024emage, liu2024towards, chen2024diffsheg, xu2024mambatalk, zhang2025echomask}. Despite these advances, several limitations remain. Directly generating high-dimensional SMPL-X motion is difficult due to complex spatial--temporal dependencies. Latent generative models alleviate this issue, but overly compact or unstructured latent spaces may weaken fine-grained facial dynamics and body-part coordination. In particular, codebook-based motion latents introduce a discrete bottleneck that may cause quantization errors and limit the continuity required for subtle expression changes, smooth gesture transitions, and coordinated holistic motion. In addition, diffusion-based generation~\cite{ho2020denoising} often requires iterative denoising, which can be inefficient and may lead to inconsistent temporal evolution for long sequences.

To address these challenges, we propose \textbf{TalkFlow}, a structured latent flow framework for speech-driven holistic motion generation. TalkFlow first learns a compact continuous motion latent space using a structure-aware autoencoder, avoiding the discretization artifacts introduced by codebook-based motion tokens. The original SMPL-X axis-angle rotations are converted to the continuous 6D rotation representation~\cite{zhou2019continuity}. Rather than splitting the body before spatial encoding, TalkFlow first encodes the full SMPL-X pose graph with a holistic GCN to preserve global body topology. At the spatial encoding stage, full-body pose and facial expression are represented as two streams; the temporal encoder subsequently separates the pose representation into upper- and lower-body branches, yielding three bottleneck streams for upper-body motion, lower-body motion, and facial expressions.

The three-stream design preserves part-specific dynamics and avoids forcing heterogeneous motion components into a single latent space. Upper-body gestures are closely related to communicative intent and hand articulation, lower-body motion provides posture and global stability, and facial expressions capture fine-grained speech-related dynamics. However, these components are not independent in natural communication. We therefore introduce multi-stage cross-stream interactions in the autoencoder, including full pose--expression interaction before temporal compression, latent-level interaction among motion streams, and pose--expression interaction after temporal decoding. These interactions are implemented through lightweight FiLM-style~\cite{perez2018film} modulation and gated residual updates. This allows the model to exchange coordination cues across body and expression representations while preserving specialized stream dynamics.

Given speech input, TalkFlow performs generation through speech-conditioned latent flow matching~\cite{lipman2022flowmatching}. Multi-level acoustic and semantic features, including HuBERT~\cite{hsu2021hubert}, Librosa~\cite{mcfee2015librosa}, FastText~\cite{bojanowski2017enriching}, and SBERT~\cite{reimers2019sentence}, are encoded into branch-specific conditions for the three motion streams. The flow model learns a continuous transport process from source latents to realistic motion latents using stream-specific Transformer branches, speech condition attention, and lightweight FiLM-gated cross-stream coordination. This design enables efficient latent-space generation while maintaining temporal coherence and holistic motion consistency.

By combining holistic graph-based motion encoding, continuous three-stream latent representation learning, and speech-conditioned latent flow matching, TalkFlow generates expressive and coordinated talking human motion. Experiments on the BEAT2 dataset~\cite{liu2024emage} demonstrate strong facial fidelity, competitive body-gesture quality, and effective speech-motion synchronization under the standard Speaker-2 evaluation protocol. Qualitative comparisons and user studies further show that TalkFlow produces more natural, speech-aligned, and perceptually preferred holistic motion compared with existing methods.

\textbf{We summarize our main contributions as follows:}
\begin{itemize}[leftmargin=1.2em, itemsep=2pt, topsep=2pt, parsep=0pt]
    \item We propose \textbf{TalkFlow}, a structured latent flow framework for speech-driven holistic motion generation that synthesizes coordinated facial expressions, upper-body gestures, and lower-body motion from speech.
    \item We introduce a \textbf{continuous structure-aware three-stream motion representation} that encodes the full-body SMPL-X pose graph and models upper-body motion, lower-body motion, and facial expressions with separate but interactive temporal streams.
    \item We formulate speech-driven holistic motion generation as \textbf{speech-conditioned latent flow matching}, using branch-specific speech conditions and a three-stream velocity backbone to achieve strong facial fidelity, competitive holistic motion performance, and improved perceptual quality on BEAT2~\cite{liu2024emage}.
\end{itemize}
\section{Related Work}
\label{sec:related work}

\subsection{Co-speech and holistic motion generation.}
Human motion generation can be conditioned on text, audio, scene context, or other control signals~\cite{zhu2023human}. Among them, speech-driven motion generation is most relevant to our work, as it requires motion to follow both acoustic dynamics and communicative semantics. Early co-speech systems relied on rule-based gesture mappings~\cite{cassell2001beat, kopp2004synthesizing, levine2010gesture, poggi2005greta}, while learning-based methods later modeled either facial animation~\cite{cudeiro2019capture, fan2022faceformer, peng2023selftalk, shen2023difftalk} or body gestures~\cite{ginosar2019learning, yoon2020speech, liu2022beat, yang2023diffusestylegesture, ao2023gesturediffuclip}. These part-specific approaches often struggle to maintain coordination across face, hands, and body. Recent holistic methods address this issue by jointly modeling multiple motion components. Habibie et al.~\cite{habibie2021learning} used a shared audio encoder with separate body-part decoders, TalkSHOW~\cite{yi2023generating} combined probabilistic gesture generation with facial animation, DiffSHEG~\cite{chen2024diffsheg} introduced directed face-to-body information flow, and EMAGE~\cite{liu2024emage} used masked gesture reconstruction and audio-conditioned synthesis. MambaTalk~\cite{xu2024mambatalk} and EchoMask~\cite{zhang2025echomask} further explored efficient temporal modeling and speech-guided masked motion generation, while SemTalk~\cite{zhang2025semtalk} emphasizes frame-level semantic cues for holistic co-speech motion generation. More recently, EchoAvatar~\cite{chen2026echoavatar} targets real-time full-body avatar animation from streaming audio. Despite these advances, preserving part-specific dynamics and holistic coordination remains challenging.

\subsection{Latent representations for holistic motion.}
Latent motion representations are widely used to reduce the complexity of high-dimensional motion generation. VQ-VAE~\cite{van2017neural} and PQ-VAE~\cite{wu2019learning, jegou2010product} provide compact discrete motion priors and have been adopted in co-speech generation. ProbTalk~\cite{liu2024towards} represents facial and upper-body motion in a shared discrete latent space, while Combo~\cite{xu2024combo} introduces part-aware latent decomposition and interaction. HoloDiff~\cite{wang2026holodiff} performs speech-conditioned diffusion in a structured latent space for holistic talking human animation. Building on structured latent modeling, TalkFlow learns continuous upper-body, lower-body, and expression streams and coordinates them through lightweight FiLM-gated interactions, preserving part-specific dynamics without relying on discrete motion tokens.

\subsection{Flow-based generative motion modeling.}
Flow-based generative models have also been explored for motion synthesis. MoGlow~\cite{henter2020moglow} introduced normalizing flows for probabilistic and controllable motion generation. More recently, Flow Matching~\cite{lipman2022flowmatching} and Rectified Flow~\cite{liu2022rectifiedflow} provide efficient continuous-time formulations for learning generative transport dynamics. In motion generation, Motion Flow Matching~\cite{hu2023motion} applies flow matching to text- and action-conditioned human motion, and MotionLab~\cite{guo2025motionlab} further studies flow-based human motion generation and editing in a unified framework. FLOAT~\cite{ki2025float} demonstrates motion latent flow matching for audio-driven talking portraits, while FlowerDance~\cite{yang2025flowerdance} explores efficient flow-based generation for music-driven dance synthesis. These works show the potential of flow-based generation, but they mainly focus on locomotion, text/action-conditioned motion, editing, dance, or talking portrait animation. TalkFlow instead applies latent flow matching to holistic 3D co-speech motion using branch-specific speech conditions and interacting upper-body, lower-body, and expression streams.
\section{Method}
\label{sec:method}

\subsection{Problem Definition}

Given a speech clip of length $N$, our goal is to generate a sequence of holistic talking human motions that are temporally aligned with the speech signal. We adopt SMPL-X~\cite{pavlakos2019expressive} as the motion representation, which models facial expressions, hand gestures, and full-body poses in a unified parametric form.

Let $A=[a_1,\dots,a_N]$ denote the input speech features. The target motion consists of articulated body poses and facial expressions. The pose sequence is denoted as $R=[r_1,\dots,r_N]$, where each pose is represented using the continuous Rot6D representation~\cite{zhou2019continuity}. Facial expressions are represented by blendshape coefficients $E=[e_1,\dots,e_N]$. Each frame contains $55\times6$ pose rotations and $100$ expression coefficients. To model heterogeneous body dynamics, we decompose the body pose into upper-body and lower-body components:
\begin{equation}
R=\{R^{u},R^{l}\},
\end{equation}
where $R^{u}$ includes upper-body and hand joints, and $R^{l}$ includes lower-body joints. Our objective is to learn a speech-conditioned generative model:
\begin{equation}
p(R^{u},R^{l},E \mid A),
\end{equation}
which produces coordinated upper-body motion, lower-body motion, and facial expressions.

\subsection{Data Preprocessing}

We preprocess the BEAT2 motion, audio, and text data into temporally aligned training clips. For motion data, we use SMPL-X pose and expression parameters, where the original axis-angle pose rotations are converted into the continuous 6D rotation representation~\cite{zhou2019continuity}. We segment each sequence into $240$-frame clips with a $60$-frame overlap. Since BEAT2 is captured at $30$ FPS, each clip corresponds to $8$ seconds of motion and is encoded into a latent sequence of length $60$ with a temporal downsampling factor of $4$. The $60$-frame overlap is used for training-clip construction and is distinct from the $8$-frame overlap used for long-sequence inference. For audio conditioning, we extract HuBERT features~\cite{hsu2021hubert} and handcrafted Librosa features~\cite{mcfee2015librosa} offline and align them with the motion clips. For textual conditioning, following MambaTalk~\cite{xu2024mambatalk}, we pre-extract word-level FastText embeddings~\cite{bojanowski2017enriching} and sentence-level SBERT embeddings~\cite{reimers2019sentence} from aligned transcripts, which are used as auxiliary conditioning signals together with audio features.

\begin{figure*}[t]
    \centering
    \includegraphics[width=1.\linewidth]{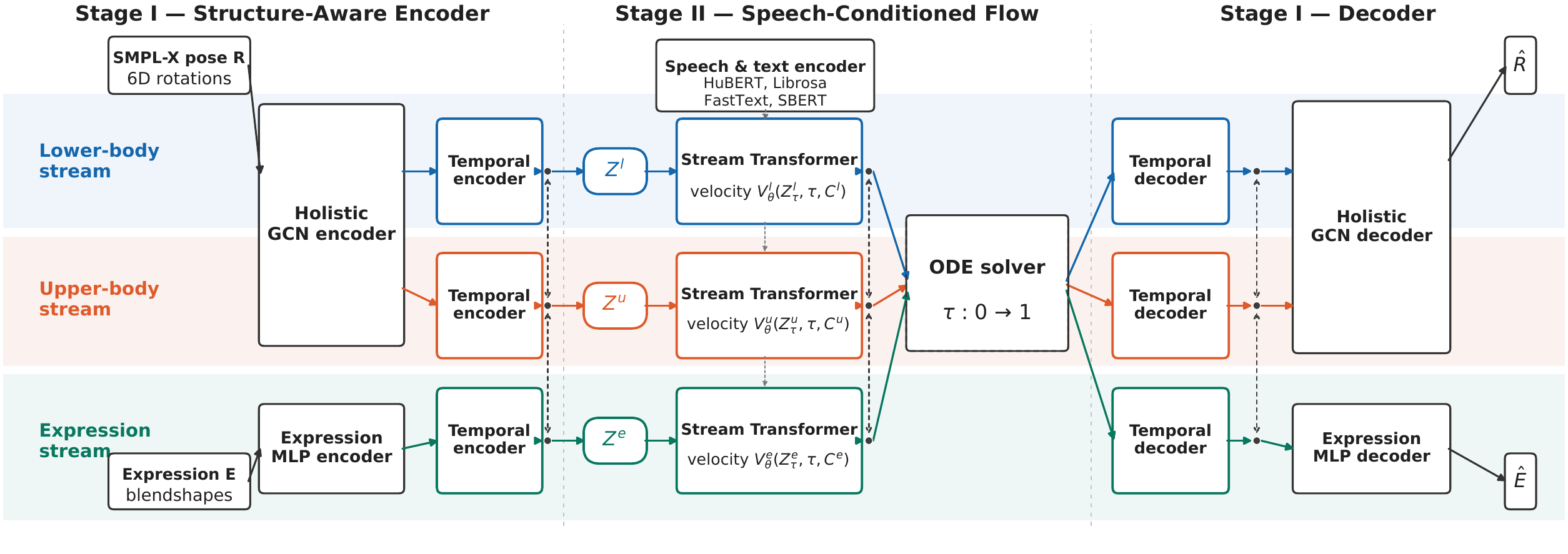}
    \caption{
    \textbf{Overall architecture of TalkFlow.}
    Stage I first encodes full-body pose and expression as two spatial branches and then forms upper-body, lower-body, and expression latent streams through temporal encoding.
    Stage II encodes speech features into branch-specific conditions and generates the three motion latents using a structured flow backbone.
    Lightweight FiLM-gated interactions coordinate upper--lower and upper--expression streams without direct lower--expression interaction.
    The frozen Stage-I decoder reconstructs the final SMPL-X motion.}
    \label{fig:pipeline}
\end{figure*}

TalkFlow follows a two-stage design, as shown in Figure~\ref{fig:pipeline}. The first stage learns a continuous structured motion space that reduces the complexity of high-dimensional SMPL-X motion while preserving part-specific dynamics and holistic coordination. The second stage performs speech-conditioned generation in this learned latent space using flow matching, allowing the flow model to operate on compact, interaction-aware motion latents.

\subsection{Stage I: Structured Three-Stream Motion Autoencoder}

We learn a structured continuous latent space for upper-body motion, lower-body motion, and facial expressions. A key design choice is to preserve full-body spatial topology before temporal decomposition. Instead of encoding upper- and lower-body poses independently from the beginning, we first encode the complete SMPL-X pose graph with a holistic graph encoder, so that global body structure and cross-limb dependencies are retained before the motion is separated into temporal streams. The graph topology follows the anatomical design of Pose2Mesh~\cite{choi2020pose2mesh}; detailed topology construction is provided in the supplementary material. We use Chebyshev graph convolutions~\cite{defferrard2016convolutional} in the pose encoder, while facial expressions are encoded by a lightweight frame-level MLP. Thus, the spatial encoding stage is initially pose--expression two-stream, whereas the compressed bottleneck is upper--lower--expression three-stream.

Given a motion sequence $(R,E)$, the spatial encoders produce
{
\begin{equation}
(H^{p},H^{p}_{g})=\mathcal{G}_{\psi}(R),
\quad
H^{e}=\mathcal{M}_{\psi}(E),
\end{equation}
}
where the graph encoder produces frame-level node features $h^{p}_{t}\in\mathbb{R}^{55\times64}$ and global pose features $h^{p}_{g,t}\in\mathbb{R}^{128}$. Their temporal stacks are denoted by $H^{p}=[h^{p}_{1},\ldots,h^{p}_{T}]$ and $H^{p}_{g}=[h^{p}_{g,1},\ldots,h^{p}_{g,T}]$, respectively.

After spatial pose--expression interaction, the updated features $\widetilde H^{p}$ and $\widetilde H^{e}$ are routed to three temporal branches:
{
\begin{equation}
Z^{u}=\mathcal{T}^{u}_{\psi}(\widetilde H^{p}),
\quad
Z^{l}=\mathcal{T}^{l}_{\psi}(\widetilde H^{p}),
\quad
Z^{e}=\mathcal{T}^{e}_{\psi}(\widetilde H^{e}).
\end{equation}
}
Here, $Z^{u}$, $Z^{l}$, and $Z^{e}$ are continuous latents for upper-body motion, lower-body motion, and facial expressions, respectively. The temporal encoder downsamples the sequence by a factor of $f$, resulting in $T'=T/f$ latent tokens. This decomposition avoids forcing heterogeneous motion components into one latent space: upper-body gestures, lower-body posture, and facial expressions exhibit different temporal dynamics and relationships to speech. The shared spatial encoder and interaction modules retain holistic structure, while the temporal branches preserve part-specific dynamics.

The decoder reconstructs upper- and lower-body pose features, merges them into a full-body representation, and outputs the original SMPL-X pose and expression:
\begin{equation}
\hat{R},\hat{E}
=
\mathcal{D}_{\psi}(Z^{u},Z^{l},Z^{e}).
\end{equation}

Although the three-stream representation preserves part-specific dynamics, independently encoding these streams would weaken holistic coordination. All interaction modules use FiLM-style modulation~\cite{perez2018film} and gated residual updates. Given a target feature $x$ and a temporally aligned conditioning feature $c$, we define
{
\begin{equation}
\begin{aligned}
h &= \phi(c),\\
(\gamma,\beta,g)
&=
\bigl(
W_{\gamma}h,\;
W_{\beta}h,\;
\sigma(W_{g}h)
\bigr),\\
\mathcal{F}(x,c)
&=
x+\alpha\,g\odot
\bigl(\gamma\odot x+\beta\bigr),\\
\mathcal{R}(x,c)
&=
x+\alpha\,g\odot W_{\Delta}h .
\end{aligned}
\label{eq:gated_interaction}
\end{equation}
}
where $\mathcal{F}$ and $\mathcal{R}$ denote the FiLM-gated and residual-gated updates, $\phi(\cdot)$ is a lightweight projection, and $\alpha$ is a learnable scalar initialized to a small value. For node-wise pose features, the modulation parameters are broadcast across the joint dimension.

At the spatial stage, expression features modulate pose-node features through $\mathcal{F}$; the updated pose nodes are pooled and softly blended with the global pose feature, which then updates the expression stream through $\mathcal{R}$. At the bottleneck, interactions are applied bidirectionally between upper--lower and upper--expression streams, while direct lower--expression interaction is omitted. After temporal decoding, a final pose--expression interaction is applied before the output decoders. This design coordinates the streams without collapsing them into a single latent representation.

The autoencoder is optimized with reconstruction and temporal consistency losses:
\begin{equation}
\mathcal{L}_{\mathrm{AE}}
=
\mathcal{L}_{\mathrm{rec}}
+
\lambda_{\mathrm{vel}}\mathcal{L}_{\mathrm{vel}}
+
\lambda_{\mathrm{acc}}\mathcal{L}_{\mathrm{acc}},
\end{equation}
where $\mathcal{L}_{\mathrm{rec}}$ supervises pose and expression reconstruction, and $\mathcal{L}_{\mathrm{vel}}$ and $\mathcal{L}_{\mathrm{acc}}$ enforce first- and second-order temporal consistency. After training, the autoencoder is frozen and used to provide structured target latents and the decoder for speech-conditioned flow generation.

\subsection{Stage II: Speech-Conditioned Latent Flow Matching}
\label{sec:flow}

Stage II generates motion by learning a speech-conditioned transport process in the structured latent space. Compared with direct motion-space generation, latent flow matching reduces the dimensionality and improves stability. Compared with codebook-based generation, it preserves a continuous motion representation and avoids quantization artifacts. The learned velocity field is integrated by an ODE solver to transport source latents to realistic motion latents.

Given the pre-extracted HuBERT, Librosa, FastText, and SBERT features, we construct branch-specific speech conditions:
\begin{equation}
C^{u},C^{l},C^{e}
=
\mathcal{C}_{\omega}(A^{h},A^{r},T^{w},T^{g}),
\end{equation}
where $A^{h}$ and $A^{r}$ denote HuBERT and Librosa audio features, and $T^{w}$ and $T^{g}$ denote word-level FastText and sentence-level SBERT features. The condition encoder produces separate conditions because facial expressions depend more strongly on phonetic and semantic content, upper-body gestures respond to prosody and communicative emphasis, and lower-body motion mainly reflects posture and stability. It uses branch-specific audio/text projections, semantic modulation, residual gating, and temporal downsampling to produce conditions aligned with the latent resolution.

Let
\begin{equation}
Z_1=\{Z^{u}_1,Z^{l}_1,Z^{e}_1\}
\end{equation}
denote target latents obtained from the frozen autoencoder. We independently sample source latents:
\begin{equation}
Z_0=\{Z^{u}_0,Z^{l}_0,Z^{e}_0\},
\qquad
Z^{s}_0\sim\mathcal{N}(0,I),
\quad
s\in\{u,l,e\}.
\end{equation}
Although ODE integration is deterministic for a fixed $Z_0$, different Gaussian source latents produce different transport trajectories and generated motions. For $\tau\sim\mathcal{U}(0,1)$, we construct the interpolated latent and target velocity:
\begin{equation}
Z_{\tau}=(1-\tau)Z_0+\tau Z_1,
\qquad
U_{\tau}=Z_1-Z_0.
\end{equation}

The flow model uses a three-stream Transformer backbone to predict velocities for the structured latent streams. Each stream performs speech-condition cross-attention with its corresponding branch-specific condition, while the motion streams exchange information through the FiLM-gated interaction defined above. We use the interaction neighborhoods $\mathcal{N}(u)=\{l,e\}$, $\mathcal{N}(l)=\{u\}$, and $\mathcal{N}(e)=\{u\}$. The velocity prediction for stream $s$ is
{
\begin{equation}
\widehat U_{\tau}^{s}
=
V_{\theta}^{s}\!\left(
Z_{\tau}^{s},\tau,C^{s};
\{Z_{\tau}^{r}\}_{r\in\mathcal{N}(s)}
\right),
\quad
s\in\{u,l,e\}.
\end{equation}
}
Thus, $\mathcal{C}=\{C^{u},C^{l},C^{e}\}$ contains three branch-specific conditions rather than one shared condition.

The latent flow matching objective is
{
\begin{equation}
\mathcal{L}_{\mathrm{FM}}
=
\sum_{s\in\{u,l,e\}}
\lambda_s
\mathbb{E}_{\tau,Z_0,Z_1}
\left[
\left\|
\widehat U_{\tau}^{s}-U_{\tau}^{s}
\right\|_2^2
\right].
\end{equation}
}
To improve compatibility between the learned latent flow and the frozen decoder, we integrate the learned field from $Z_0$ to obtain $\widetilde Z_1$, decode it through the frozen Stage-I decoder, and supervise the resulting pose and expression in the original motion space:
\begin{equation}
\mathcal{L}_{\mathrm{gen}}
=
\mathcal{L}_{\mathrm{FM}}
+
\lambda_{\mathrm{roll}}\mathcal{L}_{\mathrm{roll}}.
\end{equation}
This rollout loss constrains the integrated trajectory after decoding and improves motion-space fidelity and temporal stability.

During inference, the condition encoder first produces $\{C^{u},C^{l},C^{e}\}$ from the input speech. Source latents are then integrated from $\tau=0$ to $\tau=1$ and decoded into SMPL-X motion. We use fixed-step Euler integration with $50$ function evaluations. For arbitrary-length generation, $240$-frame windows overlap by $8$ original motion frames. The final overlap of the previous window is used as known context for the next window; overlapping predictions are linearly blended and smoothed with a Gaussian kernel of size $3$ to reduce boundary discontinuities. This stitching is confined to the overlap region and is used only for long-sequence inference.

\subsection{Training details.}
The structured autoencoder is trained first and then frozen. The flow backbone is first optimized with $\mathcal{L}_{\mathrm{FM}}$ and subsequently fine-tuned with $\mathcal{L}_{\mathrm{gen}}$. We use temporal compression $4\times$, latent dimensions $(d_z^{u},d_z^{l},d_z^{e})=(512,128,512)$, condition dimensions $(d_c^{u},d_c^{l},d_c^{e})=(256,128,384)$, a $6$-layer/$8$-head Transformer backbone with hidden dimension $384$, and interaction hidden dimension $256$. With pre-extracted speech features, Euler sampling with $50$ NFEs runs at $0.006$ s per frame on one RTX 4070 Ti SUPER. Remaining optimization and implementation details are provided in the supplementary material.

\section{Experiments}
\label{sec:experiments}

\subsection{Experimental Settings}

\subsubsection{Dataset.}
For training and evaluation, we use the BEAT2 dataset~\cite{liu2024emage}, which contains 60 hours of high-quality motion data from 25 speakers, including 12 female and 13 male speakers. The dataset consists of 1,762 sequences, with an average duration of 65.66 seconds, where speakers respond to daily inquiries. We divide the dataset into training, validation, and test sets with ratios of 85\%, 7.5\%, and 7.5\%, respectively. To ensure a fair comparison with prior work, we follow EMAGE~\cite{liu2024emage} and use the Speaker-2 subset for training, validation, and testing. This single-speaker protocol provides a controlled evaluation setting, but it also limits the range of speaker-specific gesture styles observed during training.

\subsubsection{Metrics.}
We evaluate generated body gestures using Fr\'echet Gesture Distance (FGD)~\cite{yoon2020speech}, which measures the distributional distance between generated and ground-truth gestures. We compute FGD using the feature extractor provided by the official EMAGE codebase. Following EMAGE~\cite{liu2024emage}, DIV~\cite{li2021audio2gestures} is computed as the average L1 distance between generated gesture clips, and therefore mainly reflects coordinate-level motion variation. To evaluate speech-motion synchrony, we use Beat Consistency Score (BCS)~\cite{li2021ai}, which measures the temporal alignment between gesture rhythm and low-level audio beats. For facial motion evaluation, we report vertex mean squared error (MSE) and vertex L1 difference (LVD)~\cite{yi2023generating}, which measure the discrepancy between generated and ground-truth facial vertices. Since quantitative metrics may not fully capture semantic appropriateness and perceptual naturalness, we further conduct qualitative comparisons and a user perception study.

\subsection{Quantitative Results}

\begin{table*}[t]
\centering
\caption{
\textbf{Quantitative comparison on the BEAT2 dataset.}
Following prior work, methods are grouped into facial expression generation,
body gesture generation, and holistic motion generation.
FGD is reported in $10^{-1}$ scale, BCS in $10^{-1}$ scale,
MSE in $10^{-8}$ scale, and LVD in $10^{-5}$ scale.
Bold indicates the best result, and underline indicates the second-best result.
}
\begin{tabular}{l c c c c c}
    \hline
    Methods & FGD$\downarrow$ & BCS$\uparrow$ & DIV$\uparrow$ &
    MSE$\downarrow$ & LVD$\downarrow$ \\
    \hline
    \hline

    Ground Truth & 0 & 6.86 & 13.05 & 0 & 0 \\
    \hline

    \textbf{Facial Expression Generation} & & & & & \\
    \hline

    FaceFormer~\cite{fan2022faceformer}
        & - & - & - & 7.79 & 7.59 \\
    CodeTalker~\cite{xing2023codetalker}
        & - & - & - & 8.03 & 7.77 \\
    \hline

    \textbf{Body Gesture Generation} & & & & & \\
    \hline

    S2G~\cite{ginosar2019learning}
        & 28.15 & 4.68 & 5.97 & - & - \\
    Trimodal~\cite{yoon2020speech}
        & 12.41 & 5.93 & 7.72 & - & - \\
    HA2G~\cite{liu2022learning}
        & 12.32 & 6.78 & 8.63 & - & - \\
    DisCo~\cite{liu2022disco}
        & 9.42 & 6.44 & 9.91 & - & - \\
    CaMN~\cite{liu2022beat}
        & 6.64 & 6.77 & 10.86 & - & - \\
    DiffStyleGesture~\cite{yang2023diffusestylegesture}
        & 8.81 & 7.24 & 11.49 & - & - \\
    \hline

    \textbf{Holistic Motion Generation} & & & & & \\
    \hline

    Habibie et al.~\cite{habibie2021learning}
        & 9.04 & 7.72 & 8.21 & 8.61 & 8.04 \\
    TalkSHOW~\cite{yi2023generating}
        & 6.21 & 6.95 & \textbf{13.47} & 7.79 & 7.77 \\
    EMAGE~\cite{liu2024emage}
        & 5.51 & 7.72 & 13.06 & 7.68 & 7.56 \\
    DiffSHEG~\cite{chen2024diffsheg}
        & 8.99 & 7.14 & 11.91 & 7.67 & 8.67 \\
    ProbTalk~\cite{liu2024towards}
        & 6.17 & 8.10 & 10.43 & 8.99 & 8.39 \\
    MambaTalk~\cite{xu2024mambatalk}
        & 5.37 & 7.81 & 13.05 &
        \underline{6.29} & \underline{6.90} \\
    EchoMask~\cite{zhang2025echomask}
        & \textbf{4.62} & 7.74 & \underline{13.37} &
        6.76 & 7.29 \\
    \hline
    \hline

    \textbf{TalkFlow (Ours)}
        & \underline{5.24}
        & \underline{7.84}
        & 11.48
        & \textbf{5.77}
        & \textbf{6.65} \\
    \hline
\end{tabular}
\label{tab:beat2_quantitative_results}
\end{table*}

As shown in Table~\ref{tab:beat2_quantitative_results}, TalkFlow achieves the best facial MSE and LVD, together with the second-best FGD and BCS. The improvements in MSE and LVD indicate that the structured latent representation preserves fine-grained facial dynamics. ProbTalk obtains a slightly higher BCS, whereas TalkFlow performs better on FGD, DIV, MSE, and LVD. These results also indicate that no single metric provides a complete measure of holistic generation quality. Combo~\cite{xu2024combo} is not included because its results use a different four-subject train/test protocol, and its code and pretrained models are unavailable for retraining under the common Speaker-2 setting.

For body gesture evaluation, TalkFlow achieves the second-best FGD. FGD evaluates distributional similarity in a learned gesture feature space, but does not directly measure perceptual realism, semantic appropriateness, or face--body coordination. EchoMask obtains the lowest FGD, whereas TalkFlow obtains the lowest mean realism rank in the user study. This difference suggests that feature-distribution distance and perceived motion quality capture complementary aspects of generation performance.

BCS measures low-level coincidence between audio and motion beats. Notably, the ground-truth BCS of $6.86$ is lower than that of every holistic generation method in Table~\ref{tab:beat2_quantitative_results}. Therefore, a higher BCS does not necessarily imply more natural or semantically appropriate motion and may favor frequent or regular beat responses. In contrast, the user study evaluates perceived speech-motion synchronization at a higher perceptual level, where TalkFlow obtains the lowest mean rank among the evaluated methods. BCS should thus be interpreted as a low-level rhythmic alignment metric rather than a standalone measure of overall synchronization quality.

TalkFlow obtains a lower quantitative DIV than several baselines. Since DIV is computed from L1 distances between generated gesture clips, it mainly reflects coordinate-level motion variation rather than high-level diversity of gesture patterns, expressive modes, or talking styles. This distinction is reflected in the user study: MambaTalk obtains the lowest mean diversity rank, while TalkFlow ranks second, despite their lower quantitative DIV than TalkSHOW and EchoMask. Flow matching is not inherently deterministic, because different Gaussian source latents produce different generation trajectories. However, the current model uses a simple white-noise prior without explicit talking-style conditioning, and the Speaker-2 protocol contains limited style variation. Style-conditioned priors and multi-speaker training may therefore further improve controllable diversity.

\begin{figure*}
    \centering
    \includegraphics[width=.95\linewidth]{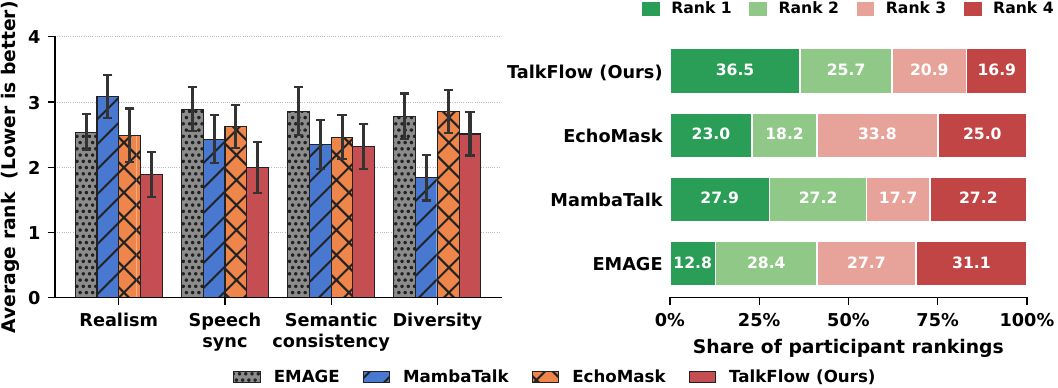}
\caption{
  \textbf{Perceptual user study.}
  $N{=}37$ participants ranked TalkFlow against three baselines on four explicitly labeled criteria.
  \textbf{Left:} mean rank per criterion, with error bars denoting one standard deviation across participants; lower is better.
  \textbf{Right:} percentage shares of Rank-1 to Rank-4 votes for each method.
  TalkFlow obtains the lowest mean rank in realism, semantic consistency, and speech-motion synchronization, and the largest overall share of first-place votes.
}
    \label{fig:user_study}
\end{figure*}

\subsection{Ablation Study}

\begin{table}[t]
\centering
\caption{
\textbf{Ablation study on the BEAT2 dataset.}
``Single latent'', ``Dual latent'', and ``Triple latent'' denote the number of latent streams.
``Int.'' denotes cross-stream interaction, and ``Rollout Rec.'' denotes training with ODE rollout reconstruction loss.
}
\setlength{\tabcolsep}{3.5pt}
\begin{tabular}{l c c c c c}
    \hline
    Variant & FGD$\downarrow$ & BCS$\uparrow$ & DIV$\uparrow$ &
    MSE$\downarrow$ & LVD$\downarrow$ \\
    \hline
    \hline

    Single latent
        & 13.21 & 7.45 & 8.67 & 7.20 & 7.76 \\
    \hline

    Dual latent
        & 11.48 & 7.72 & 9.07 & 6.66 & 7.16 \\
    \hline

    \textbf{Triple latent} & & & & & \\
    \hline

    w/o Int.
        & 6.40 & 7.80 & 10.94 & 6.11 & 6.79 \\

    w/o Flow Int.
        & 5.30 & 7.73 & 11.20 & 6.09 & 6.75 \\

    w/o AE Int.
        & 5.40 & 7.72 & 11.01 & 6.30 & 7.09 \\

    w/o GCN
        & 5.49 & 7.76 & 10.82 & 6.16 & 6.94 \\

    w/o Rollout Rec.
        & 9.46 & 7.75 & 10.84 & 6.44 & 6.97 \\
    \hline
    \hline

    \textbf{Full}
        & \textbf{5.24}
        & \textbf{7.84}
        & \textbf{11.48}
        & \textbf{5.77}
        & \textbf{6.65} \\
    \hline
\end{tabular}
\label{tab:beat2_ablation_results}
\end{table}

Table~\ref{tab:beat2_ablation_results} validates the contribution of the main components in TalkFlow. The single-latent variant compresses all motion components into one shared latent space. Separating pose and expression in the dual-latent variant improves facial MSE/LVD from $7.20/7.76$ to $6.66/7.16$, while also improving FGD, BCS, and DIV. The triple-latent variants further decompose full-body pose into upper- and lower-body streams, producing a substantial reduction in FGD and confirming the benefit of preserving heterogeneous part-specific dynamics.

Introducing autoencoder-stage interaction further improves both body and facial metrics, indicating that coordination during representation learning benefits holistic motion generation. The full model achieves the best results across all ablation metrics, showing that graph-based pose encoding, autoencoder and flow-stage interactions, and ODE rollout reconstruction are complementary. In particular, removing rollout reconstruction substantially degrades FGD, confirming that motion-space rollout supervision improves compatibility between the learned latent trajectory and the frozen decoder.

\subsection{User Perception Study}

Following the user study protocol of EchoMask~\cite{zhang2025echomask}, we conduct a perceptual evaluation to compare the visual quality and speech-motion consistency of different methods. We recruit 37 participants with normal or corrected-to-normal vision, normal hearing, and sufficient English proficiency to understand the evaluated speech clips. Participants are asked to rank generated videos from different methods according to four perceptual criteria: realism, semantic consistency, speech-motion synchronization, and motion diversity. We report the average rank for each method, where a lower rank indicates stronger user preference.

As shown in Fig.~\ref{fig:user_study}, TalkFlow obtains the lowest mean rank in realism, semantic consistency, and speech-motion synchronization. These perceptual results complement rather than replace the quantitative metrics. Although EchoMask obtains a lower FGD in Table~\ref{tab:beat2_quantitative_results}, TalkFlow receives a lower mean realism rank, illustrating that FGD does not fully capture motion naturalness, temporal stability, and holistic coordination.

For speech-motion synchronization, TalkFlow obtains the lowest mean user-study rank among the evaluated methods, while its quantitative BCS is second-best. This difference is consistent with the distinct aspects captured by the two evaluations: BCS measures low-level audio-beat coincidence, whereas participants evaluate whether generated gestures and facial motion appear appropriately synchronized with the speech as a whole. The comparatively low ground-truth BCS further shows that maximizing beat coincidence is not equivalent to maximizing perceived synchronization.

For motion diversity, MambaTalk obtains the lowest mean rank and TalkFlow ranks second. This differs from quantitative DIV, where both methods score below TalkSHOW and EchoMask. DIV reflects coordinate-level variation and therefore does not necessarily capture high-level variation in gesture patterns, expressive modes, or talking styles. The user-study results suggest that TalkFlow retains perceptually meaningful motion variation despite its lower coordinate-level DIV.

Participant comments further support these observations. EMAGE often produces body movements that are too calm and do not fully match the agitated tone of the speech; however, its conservative motion also avoids severe torso deformation. MambaTalk exhibits the largest range of body movement and often matches the angry tone well, but participants also observe occasional unnatural jitters and sudden full-body twists. EchoMask sometimes produces repetitive arm postures, and its wrists may appear unnatural. In contrast, TalkFlow is generally perceived as the most balanced method, with smoother pose transitions, more stable body motion, and better synchronization with the speech tempo.

\subsection{Qualitative Results}

\begin{figure}[t]
    \centering
    \setlength{\tabcolsep}{1pt}
    \renewcommand{\arraystretch}{0.85}
    \setlength{\fboxrule}{0.8pt}
    \setlength{\fboxsep}{0pt}

    \newcommand{\qualimg}[1]{%
        \includegraphics[
            width=0.145\linewidth,
            height=0.075\textheight,
            keepaspectratio
        ]{#1}
    }

    \newcommand{\qualimgred}[1]{%
        \fcolorbox{red}{white}{\qualimg{#1}}%
    }

    \scriptsize
    \begin{tabular}{c c c c c c}
        & \textbf{am \textcolor{red}{not}}
        & \textbf{colle\textcolor{red}{ge}}
        & \textbf{in \textcolor{red}{dan}gerous}
        & \textbf{inter\textcolor{red}{net}}
        & \textbf{\textcolor{red}{kind} of} \\

        \textbf{Ground Truth} &
        \qualimg{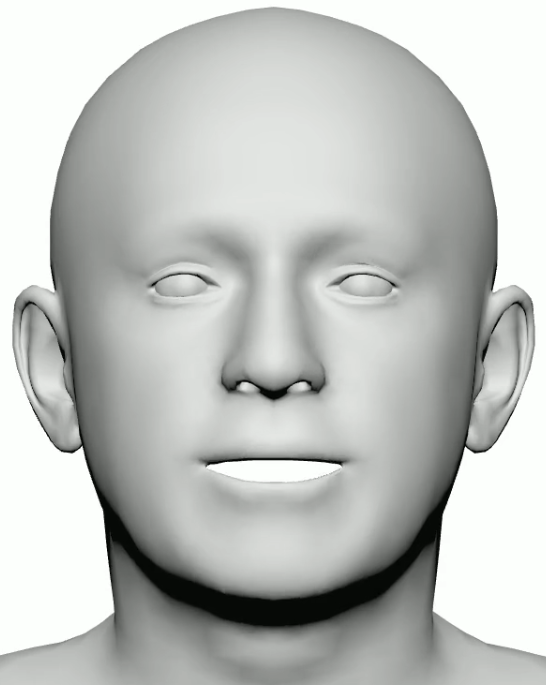} &
        \qualimg{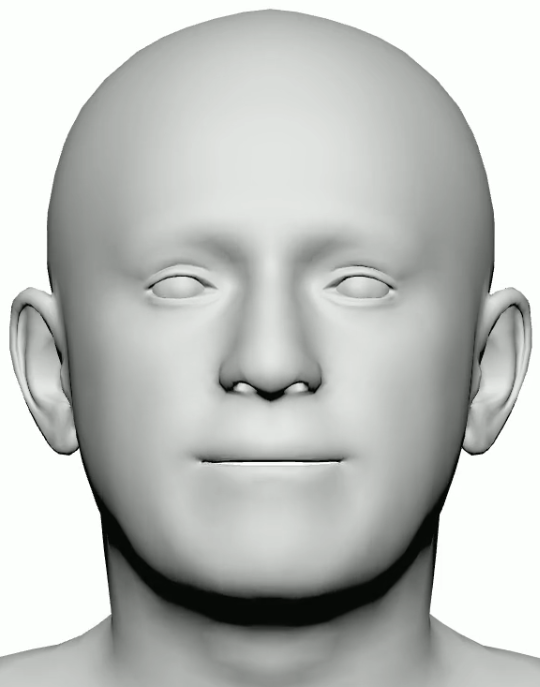} &
        \qualimg{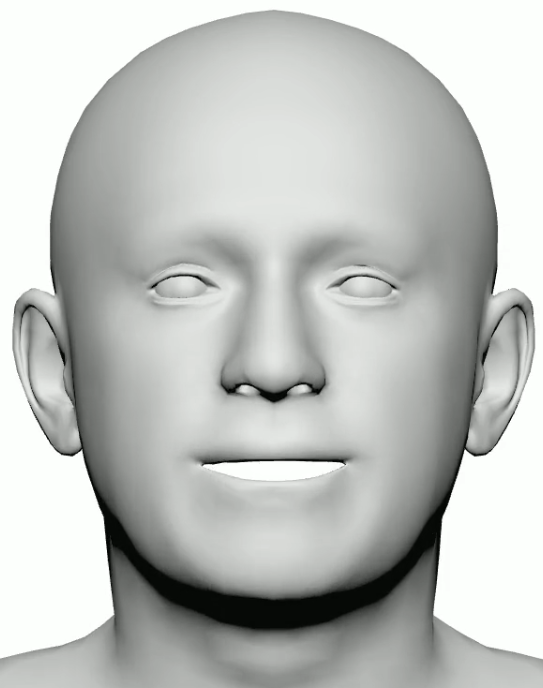} &
        \qualimg{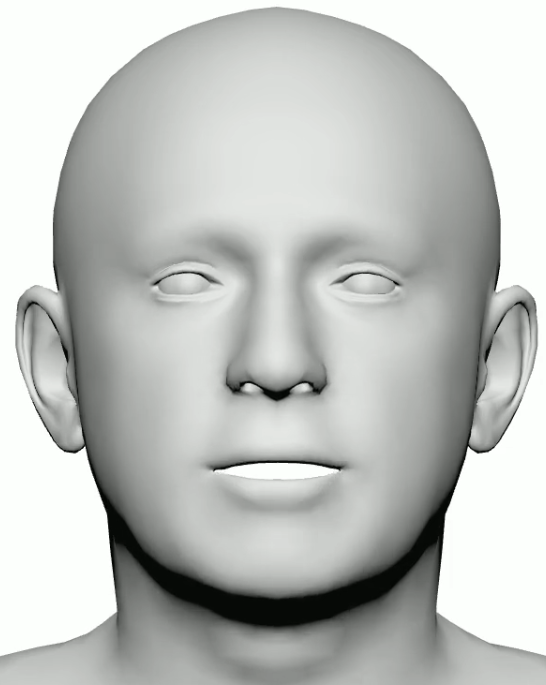} &
        \qualimg{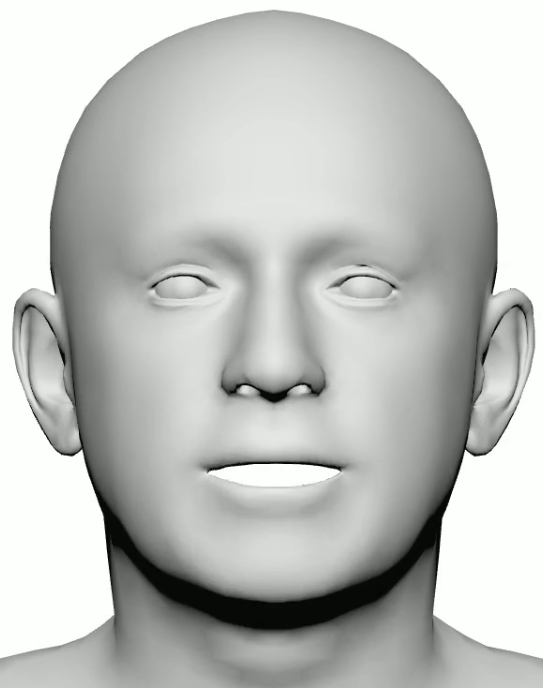} \\

        \textbf{TalkFlow (Ours)} &
        \qualimg{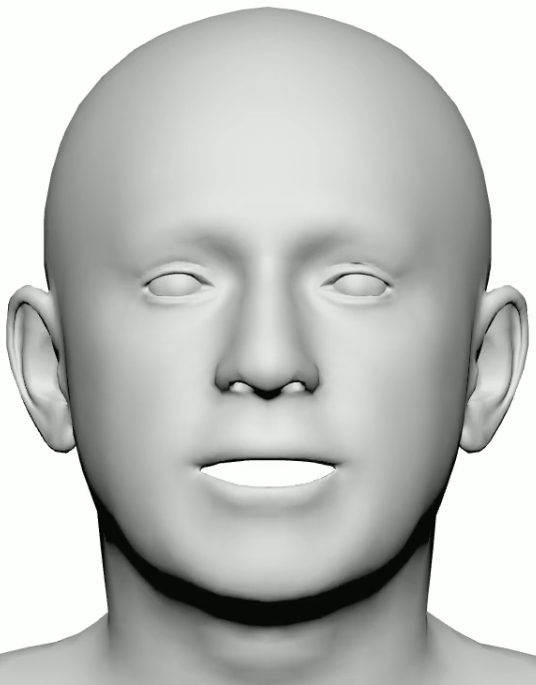} &
        \qualimg{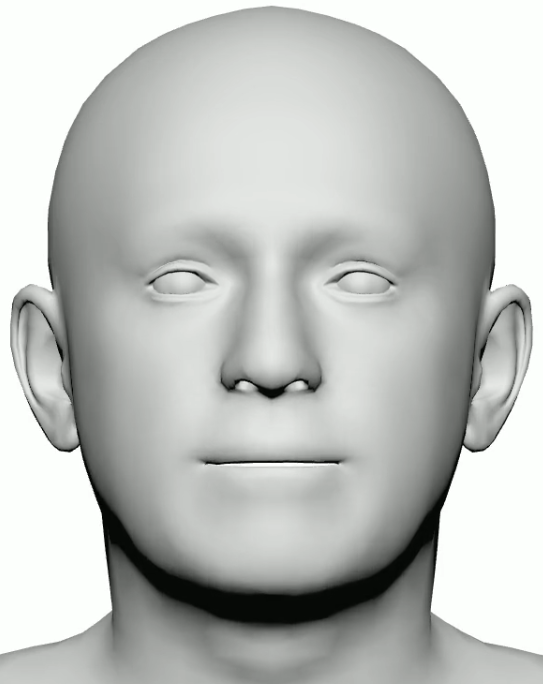} &
        \qualimg{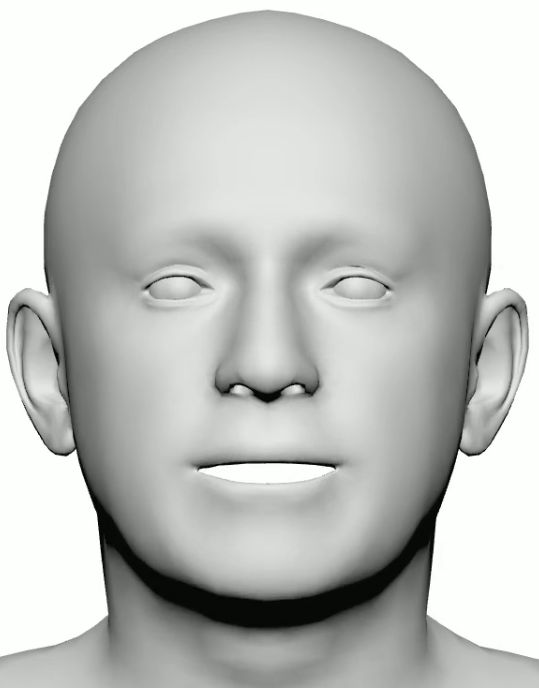} &
        \qualimg{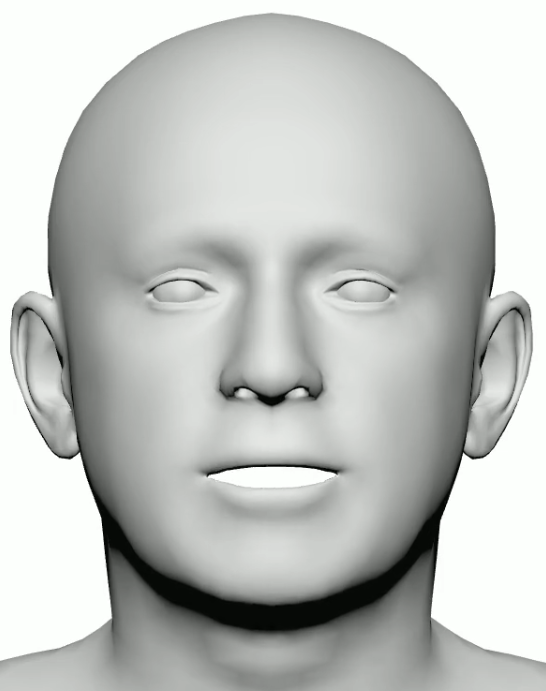} &
        \qualimg{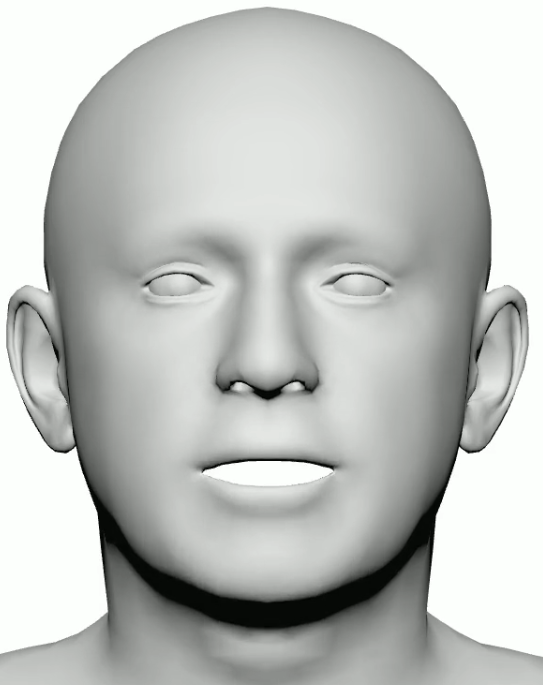} \\

        \textbf{EchoMask} &
        \qualimgred{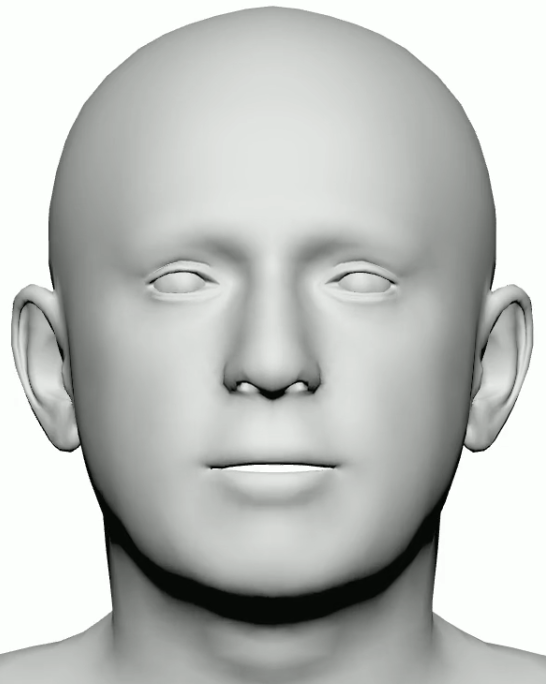} &
        \qualimgred{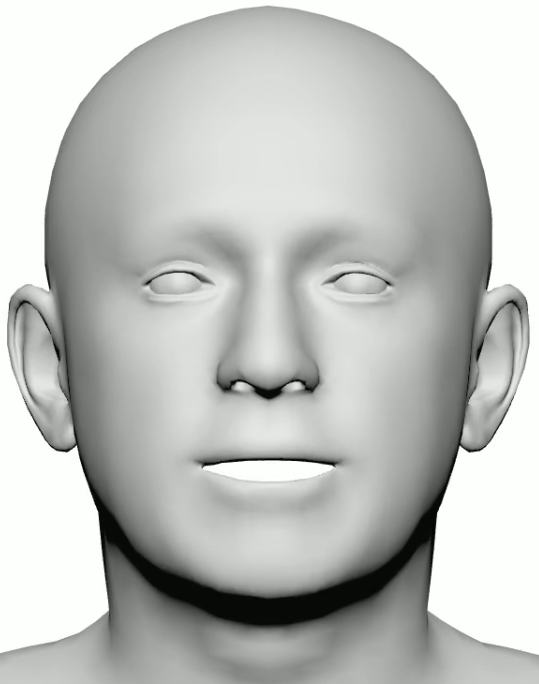} &
        \qualimgred{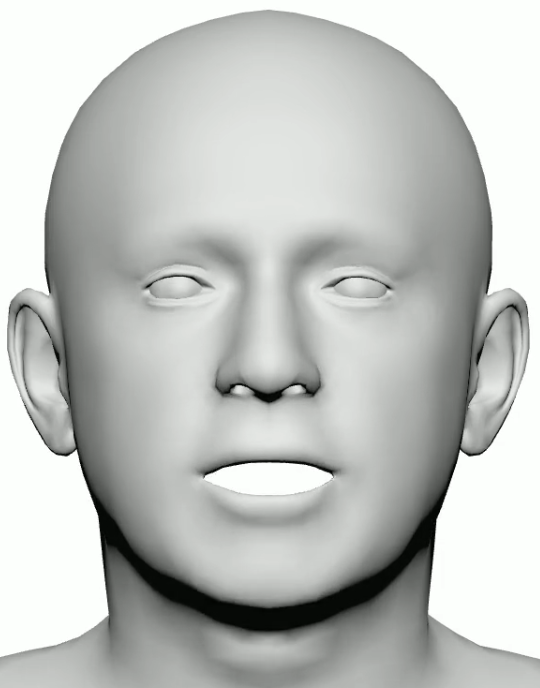} &
        \qualimg{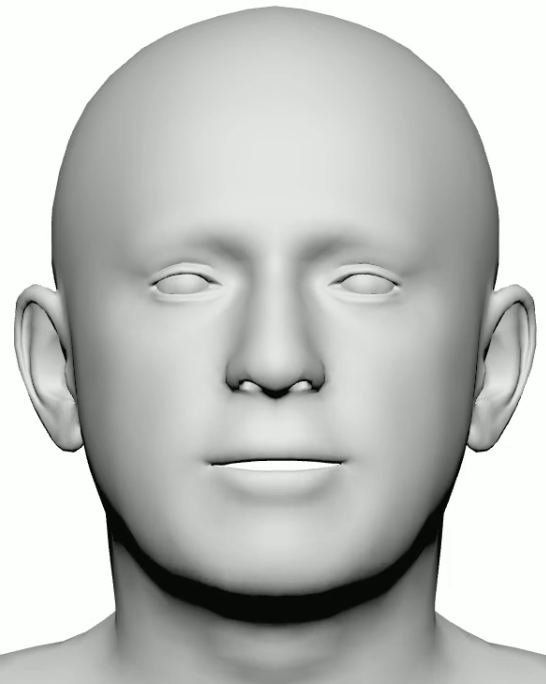} &
        \qualimg{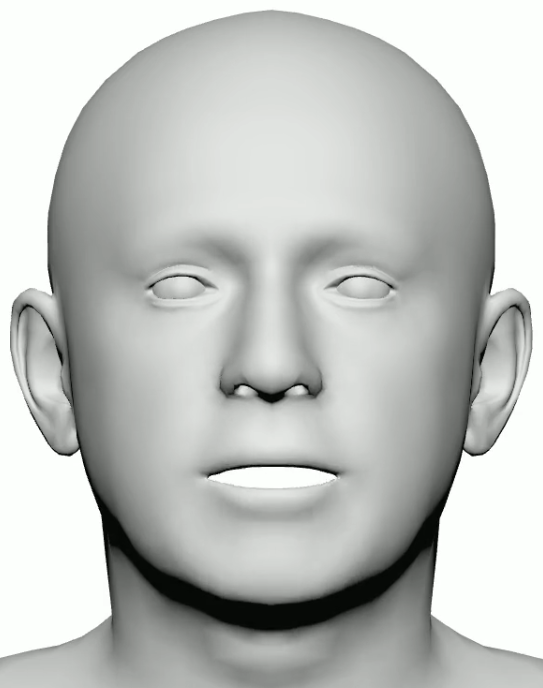} \\

        \textbf{EMAGE} &
        \qualimg{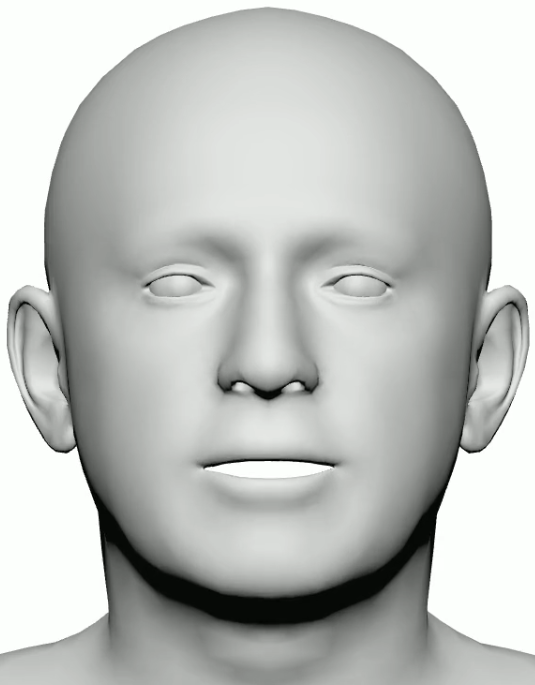} &
        \qualimgred{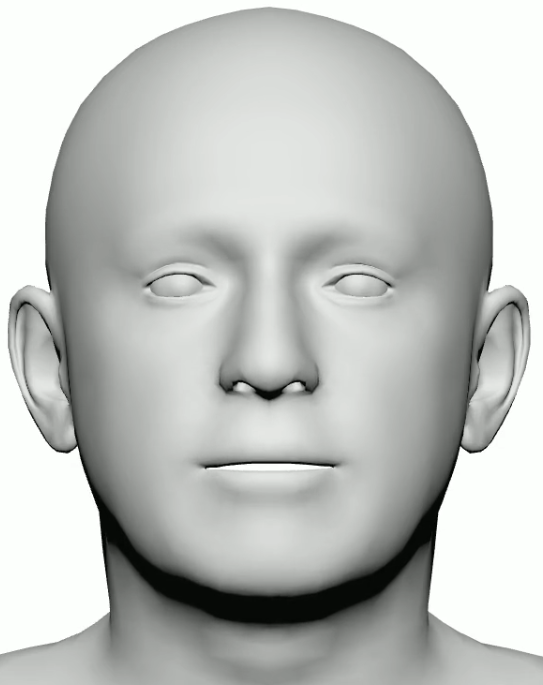} &
        \qualimgred{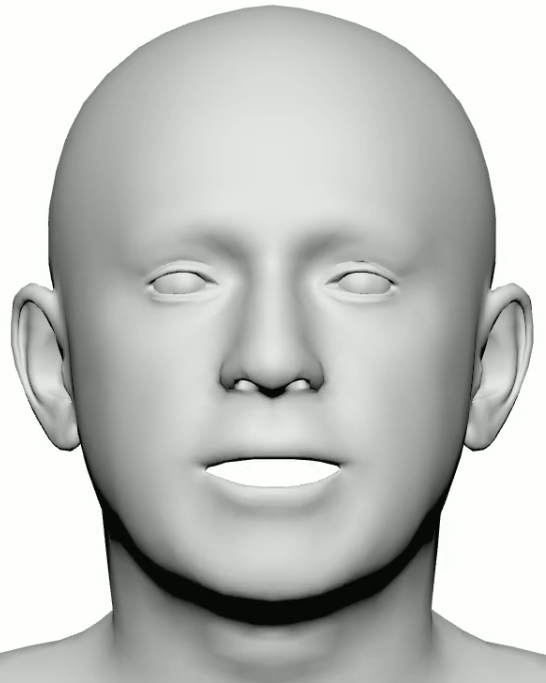} &
        \qualimgred{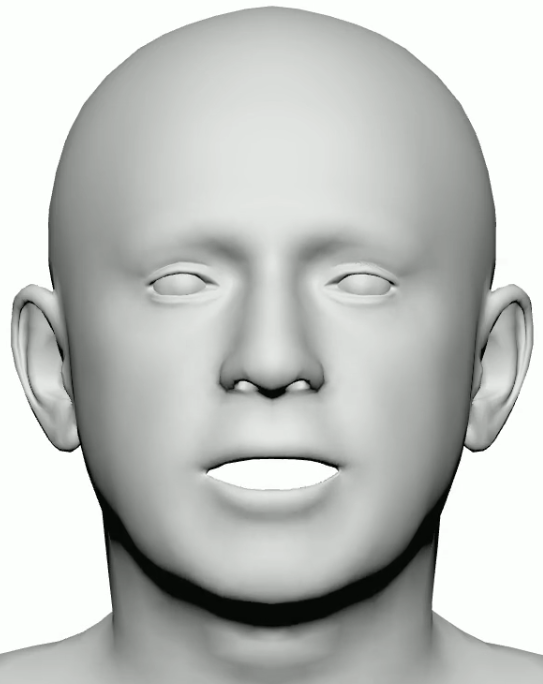} &
        \qualimg{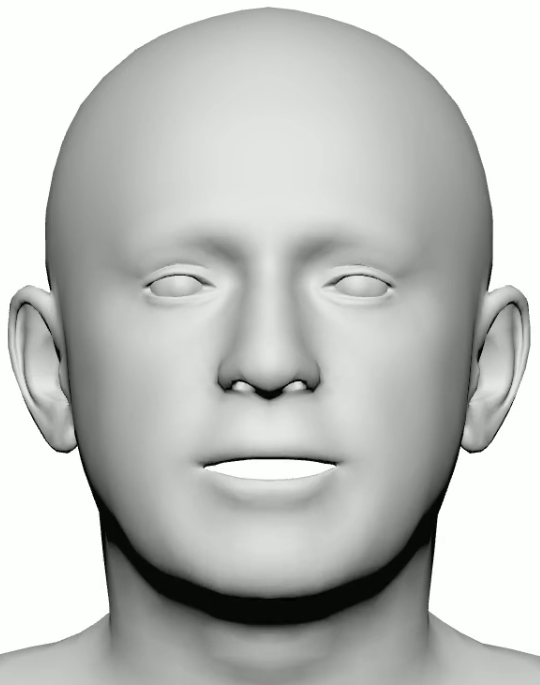} \\

        \textbf{MambaTalk} &
        \qualimg{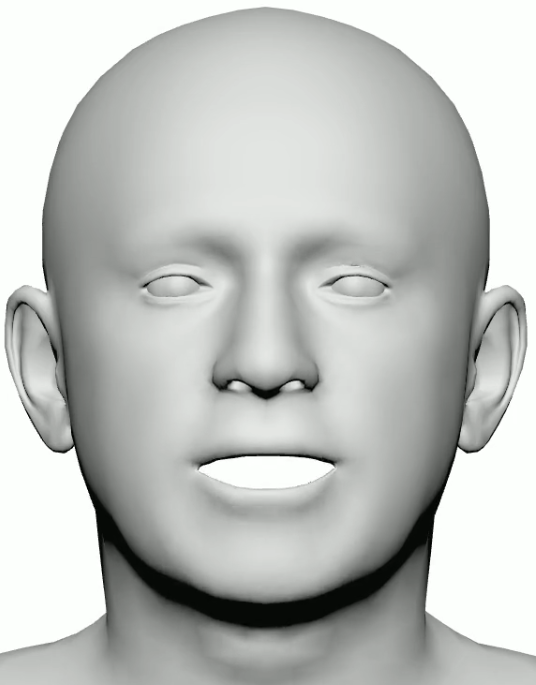} &
        \qualimgred{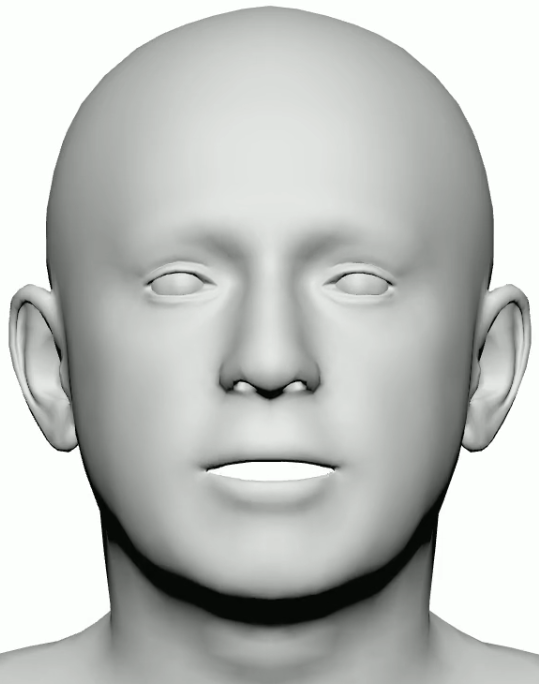} &
        \qualimg{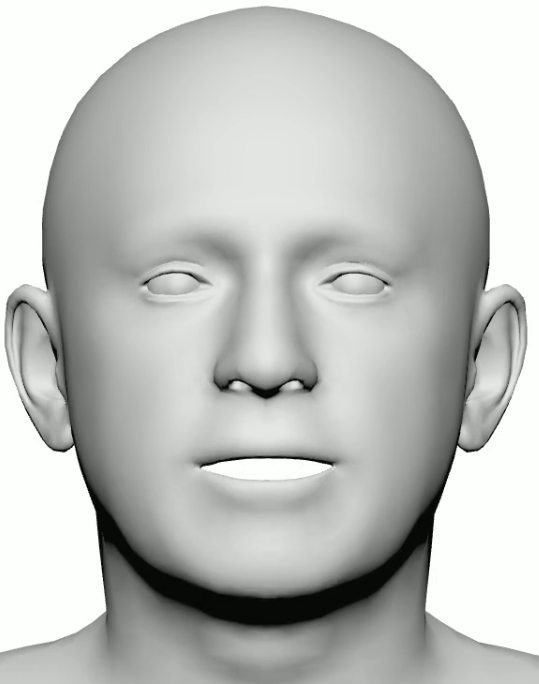} &
        \qualimg{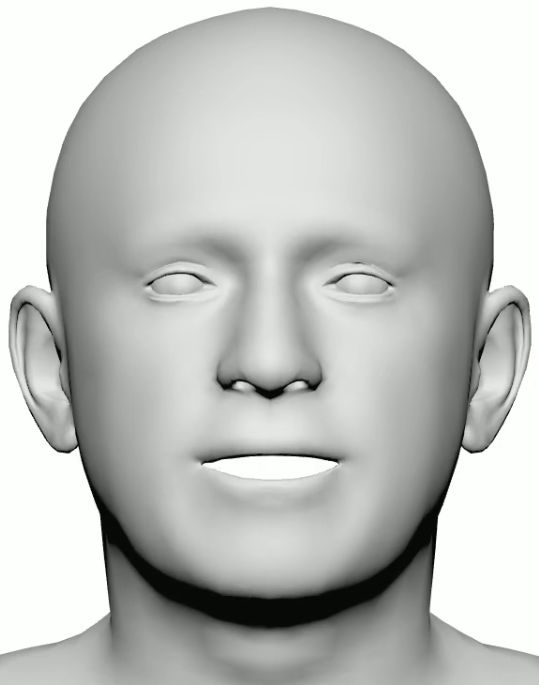} &
        \qualimgred{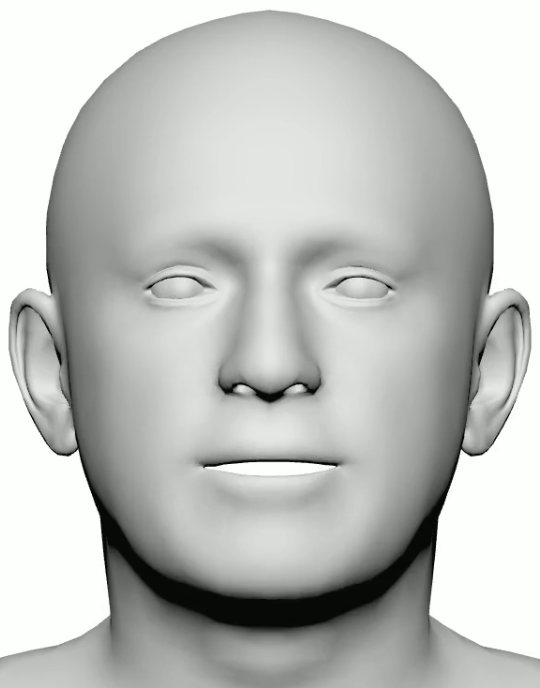} \\
    \end{tabular}

    \caption{
    Facial qualitative comparison on selected spoken words.
    Each column corresponds to a spoken word, with the highlighted syllable shown in red.
    Red boxes indicate evident lip-shape errors relative to the ground truth, such as an open mouth versus a closed mouth.
    TalkFlow generally remains closest to the ground-truth mouth shape across the selected comparisons.
    }
    \label{fig:qual_head_single}
\end{figure}

{
We provide qualitative comparisons with representative holistic co-speech motion generation methods in the final two figure-only pages to allow larger rendered frames and clearer visual inspection. We first compare facial motion at selected spoken words in Fig.~\ref{fig:qual_head_single}. Since BEAT2 facial motion is generally close to neutral and SMPL-X provides a relatively sparse facial representation, differences in static renderings are mainly visible through lip opening/closing and mouth-motion timing. TalkFlow generally produces mouth shapes closest to the ground truth, while several baseline results contain clear lip-shape errors, such as an open mouth when the ground-truth mouth is closed. Red boxes mark only these evident errors; unboxed results are not assigned a binary correct/incorrect label.
}

\begin{figure*}[t]
    \centering
    \setlength{\tabcolsep}{1pt}
    \renewcommand{\arraystretch}{0.9}

    \newlength{\qualheight}
    \setlength{\qualheight}{0.15\textheight}

    \newlength{\methodwidth}
    \setlength{\methodwidth}{0.12\textwidth}

    \newlength{\imgcellwidth}
    \setlength{\imgcellwidth}{0.123\textwidth}


    \newcommand{\methodcell}[1]{%
        \makebox[\methodwidth][c]{%
            \raisebox{0.06\textheight}[0pt][0pt]{\textbf{#1}}%
        }%
    }

    \newcommand{\imgcell}[1]{%
        \makebox[\imgcellwidth][c]{%
            \includegraphics[
                height=\qualheight,
                keepaspectratio
            ]{#1}%
        }%
    }

    \newcommand{\wordcell}[1]{%
        \makebox[\imgcellwidth][c]{%
            \begin{minipage}{\imgcellwidth}
                \centering
                \small\textbf{#1}
            \end{minipage}%
        }%
    }

    \makebox[\textwidth][c]{%
    \begin{tabular}{@{} c c c c c c c c @{}}

        \methodcell{Ground Truth} &
        \imgcell{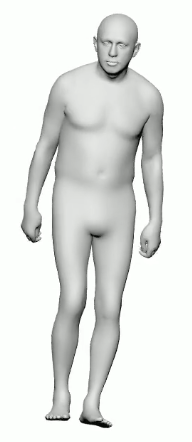} &
        \imgcell{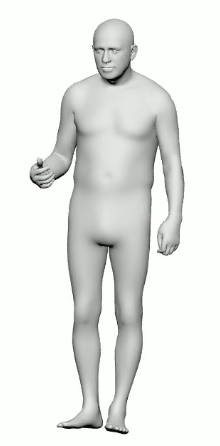} &
        \imgcell{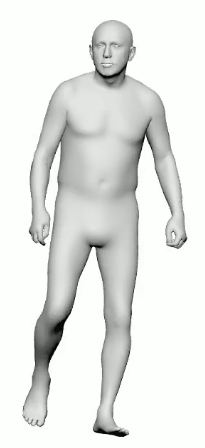} &
        \imgcell{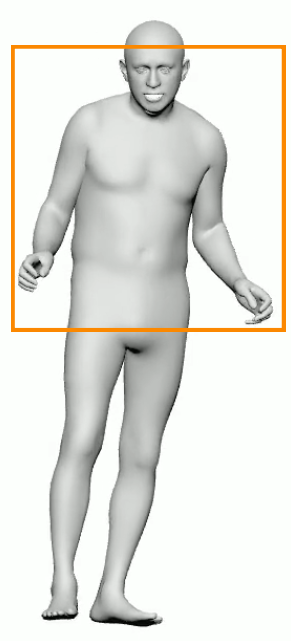} &
        \imgcell{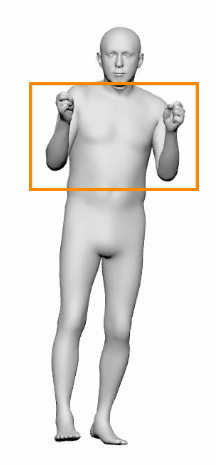}&
        \imgcell{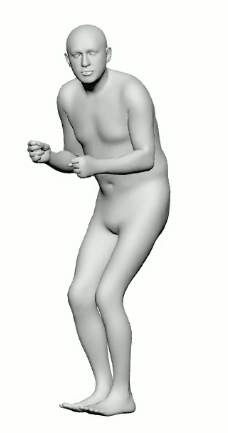}&
        \imgcell{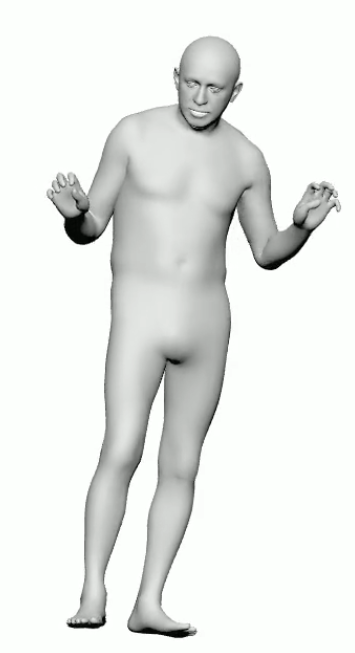}\\[-0.5mm]

        &
        \wordcell{the \textcolor{blue}{hospital} ...}&
        \wordcell{\textcolor{blue}{talking} loudly...}&
        \wordcell{terrible ...}&
        \wordcell{commit \textcolor{blue}{suicide} ...} &
        \wordcell{\textcolor{blue}{watching} ...}&
        \wordcell{so \textcolor{blue}{scared} ...}&
        \wordcell{don't \textcolor{blue}{share} ...} \\[2mm]

        \methodcell{TalkFlow} &
        \imgcell{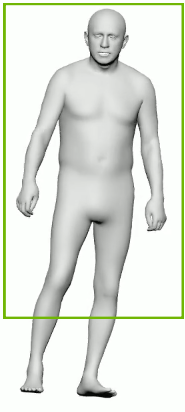} &
        \imgcell{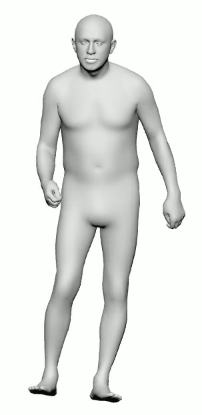} &
        \imgcell{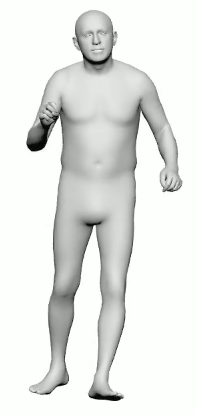} &
        \imgcell{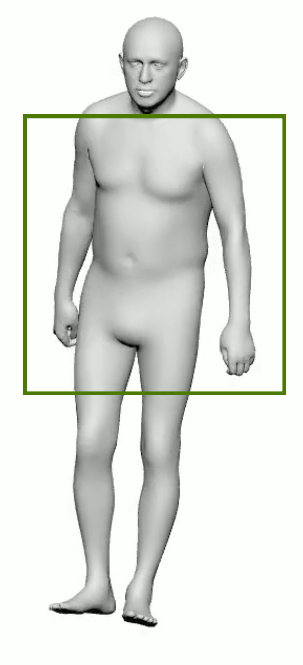} &
        \imgcell{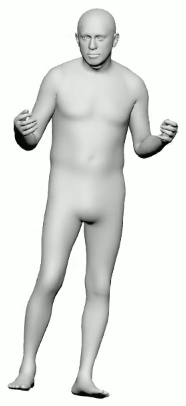}&
        \imgcell{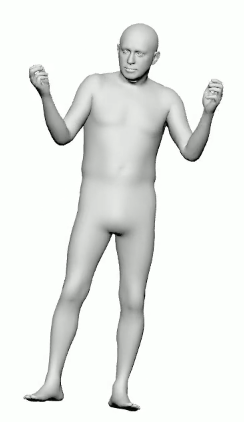}&
        \imgcell{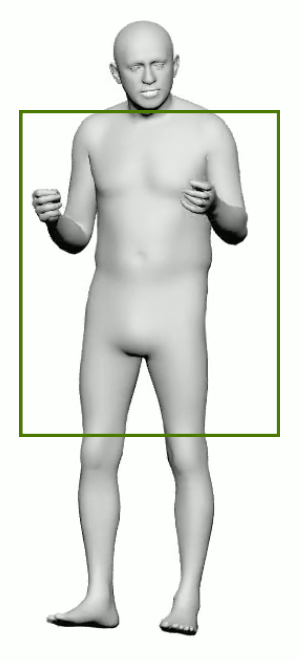}\\[-0.5mm]

        \methodcell{EchoMask} &
        \imgcell{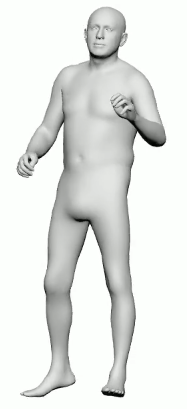} &
        \imgcell{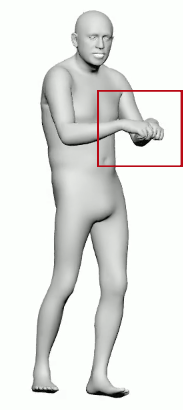} &
        \imgcell{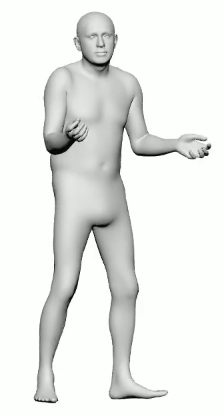} &
        \imgcell{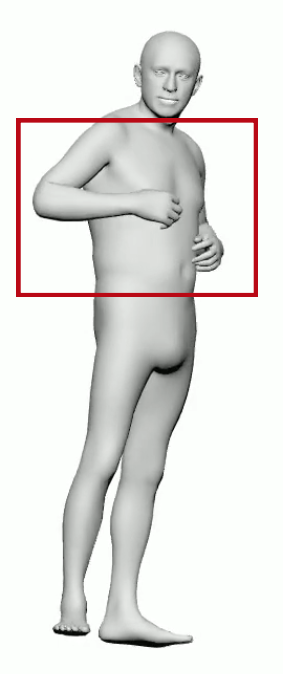} &
        \imgcell{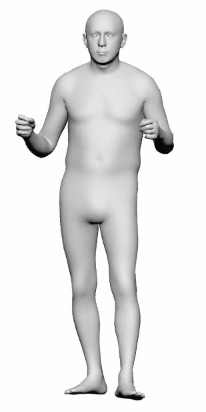}&
        \imgcell{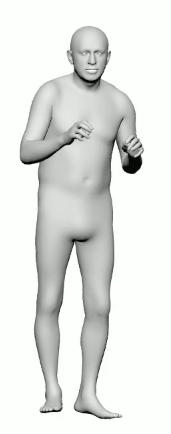}&
        \imgcell{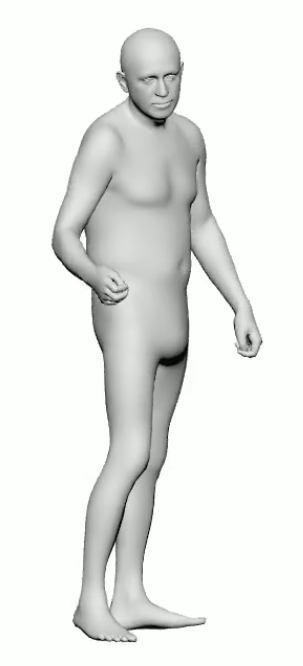}\\[-0.5mm]

        \methodcell{EMAGE} &
        \imgcell{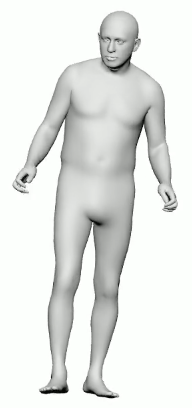} &
        \imgcell{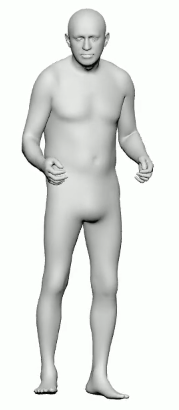} &
        \imgcell{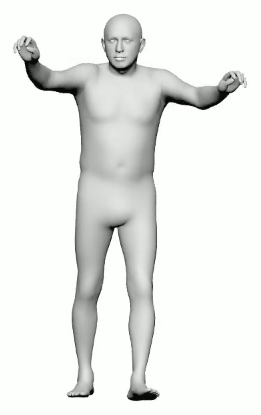} &
        \imgcell{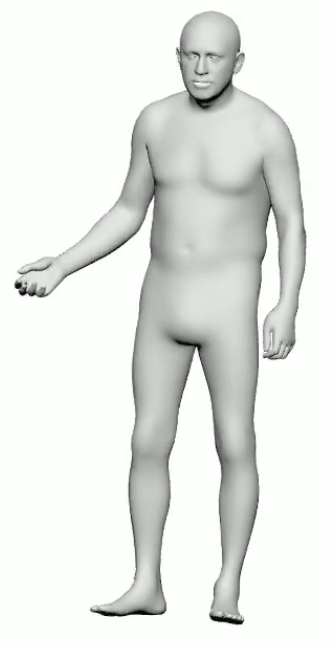} &
        \imgcell{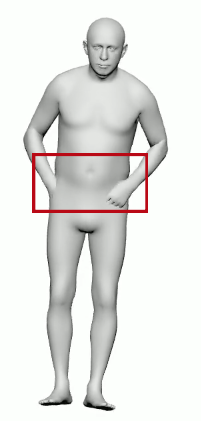}&
        \imgcell{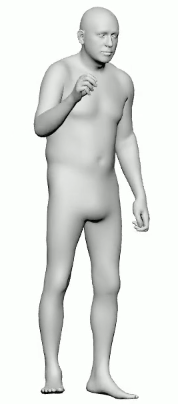}&
        \imgcell{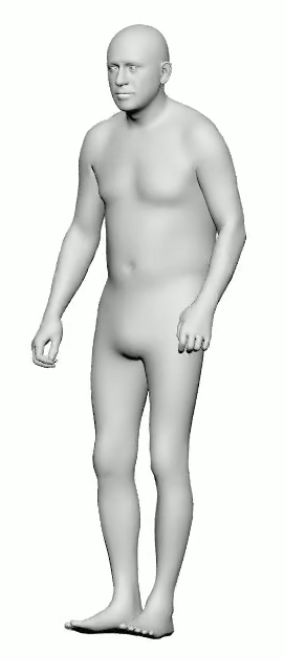}\\[-0.5mm]

        \methodcell{MambaTalk} &
        \imgcell{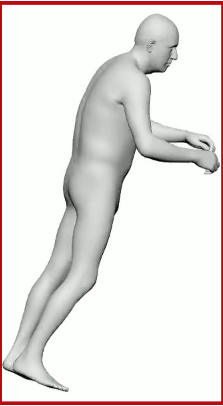} &
        \imgcell{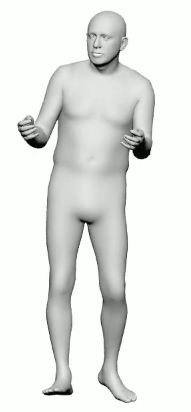} &
        \imgcell{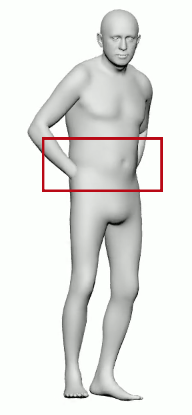} &
        \imgcell{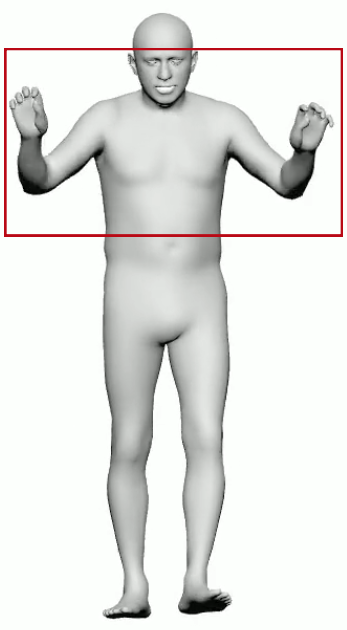} &
        \imgcell{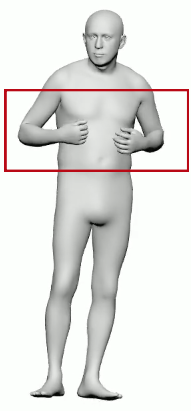}&
        \imgcell{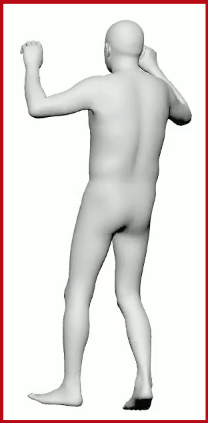}&
        \imgcell{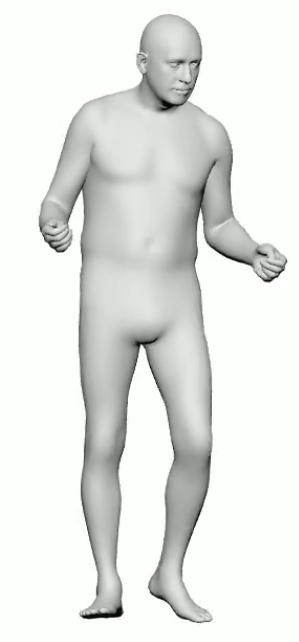}\\

    \end{tabular}%
    }

    \caption{ \textbf{Qualitative comparison on representative word-level speech contents.} The first row shows ground-truth motions and the corresponding speech contents, followed by results from different methods. Orange boxes highlight representative ground-truth motion regions, green boxes indicate generated motions closest to the corresponding references, and red boxes mark visible artifacts or semantically inconsistent motions. For EchoMask on ``talking loudly,'' the two hands intersect. For MambaTalk on ``terrible'' and EMAGE on ``watching,'' the fists intersect the torso; MambaTalk also exhibits abrupt whole-body rotation and imbalance. }
    \label{fig:qualitative_comparison_1}
\end{figure*}

\begin{figure*}[t]
    \centering
    \setlength{\tabcolsep}{1pt}
    \renewcommand{\arraystretch}{0.9}

    \setlength{\qualheight}{0.15\textheight}

    \setlength{\methodwidth}{0.12\textwidth}

    \setlength{\imgcellwidth}{0.123\textwidth}


    \newcommand{\methodcell}[1]{%
        \makebox[\methodwidth][c]{%
            \raisebox{0.06\textheight}[0pt][0pt]{\textbf{#1}}%
        }%
    }

    \newcommand{\imgcell}[1]{%
        \makebox[\imgcellwidth][c]{%
            \includegraphics[
                height=\qualheight,
                keepaspectratio
            ]{#1}%
        }%
    }

    \newcommand{\wordcell}[1]{%
        \makebox[\imgcellwidth][c]{%
            \begin{minipage}{\imgcellwidth}
                \centering
                \small\textbf{#1}
            \end{minipage}%
        }%
    }

    \makebox[\textwidth][c]{%
    \begin{tabular}{@{} c c c c c c c c @{}}

        \methodcell{Ground Truth} &
        \imgcell{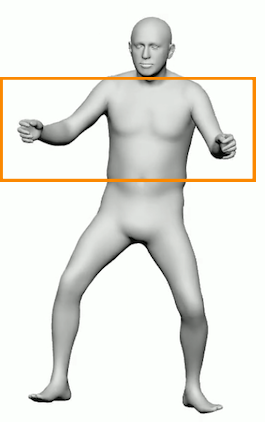} &
        \imgcell{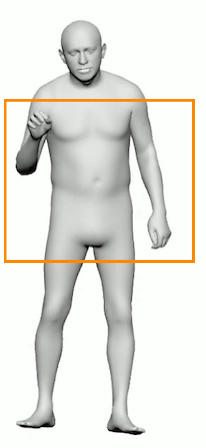} &
        \imgcell{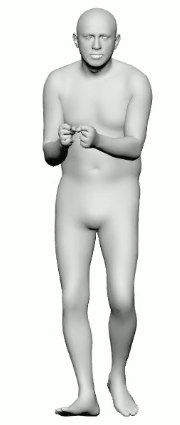} &
        \imgcell{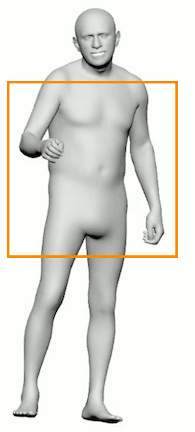} &
        \imgcell{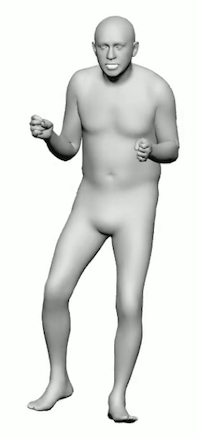}&
        \imgcell{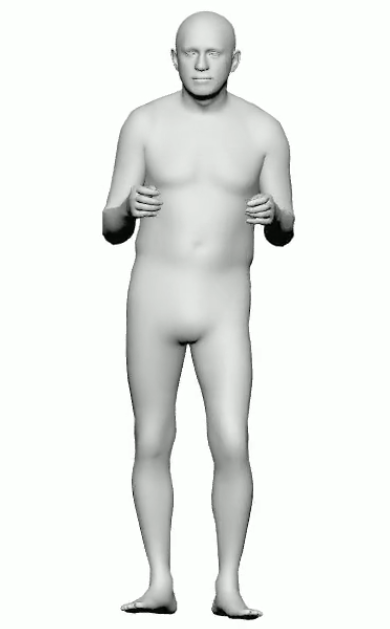}&
        \imgcell{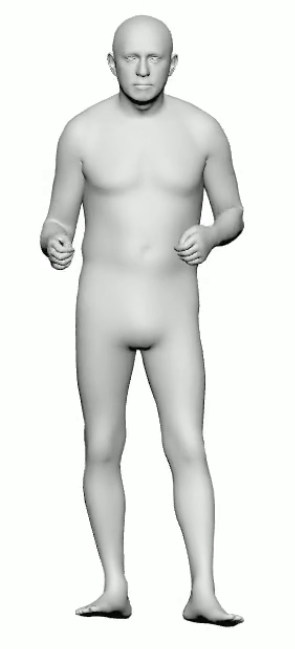}\\[-0.5mm]

        &
        \wordcell{wow ...}&
        \wordcell{especially...}&
        \wordcell{in the \textcolor{blue}{street} ...}&
        \wordcell{no ...} &
        \wordcell{police \textcolor{blue}{shot} ...}&
        \wordcell{two \textcolor{blue}{months} ...}&
        \wordcell{it is \textcolor{blue}{delayed} ...}\\[2mm]

        \methodcell{TalkFlow} &
        \imgcell{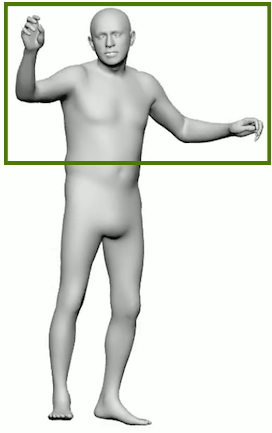} &
        \imgcell{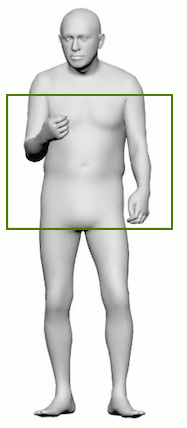} &
        \imgcell{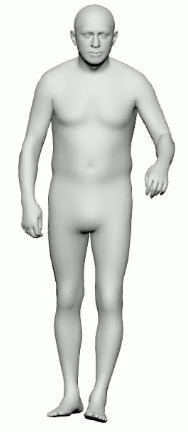} &
        \imgcell{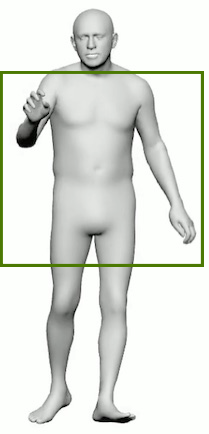} &
        \imgcell{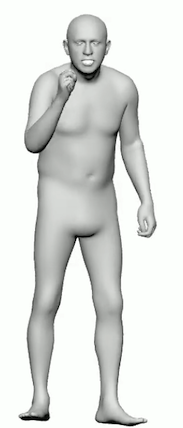}&
        \imgcell{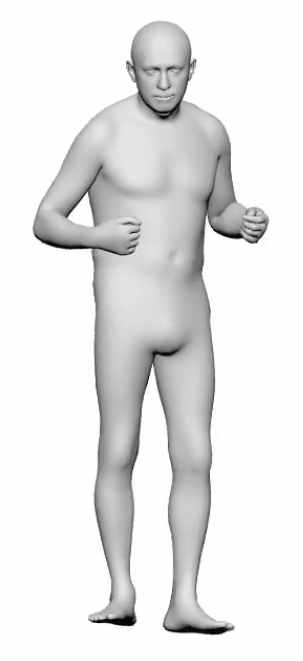}&
        \imgcell{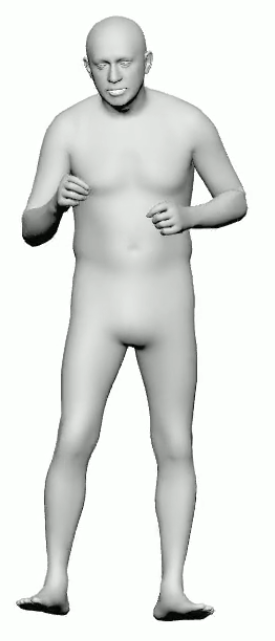}\\[-0.5mm]

        \methodcell{EchoMask} &
        \imgcell{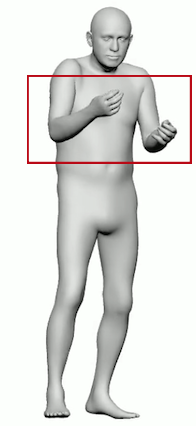} &
        \imgcell{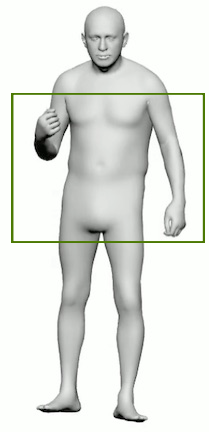} &
        \imgcell{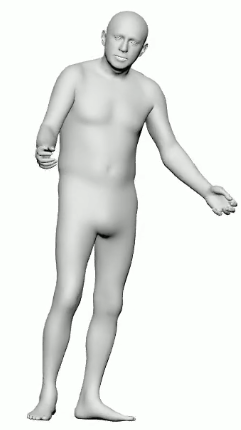} &
        \imgcell{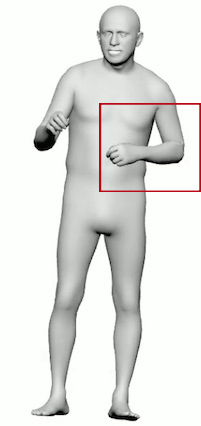} &
        \imgcell{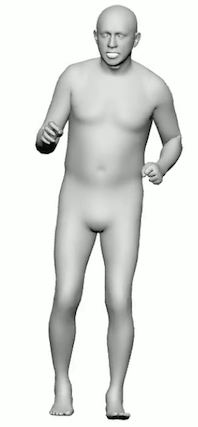}&
        \imgcell{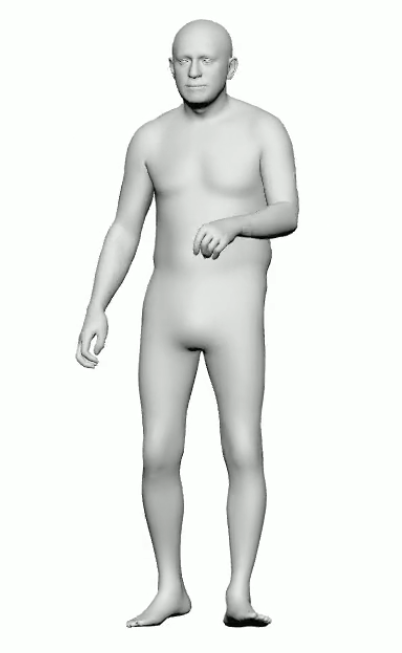}&
        \imgcell{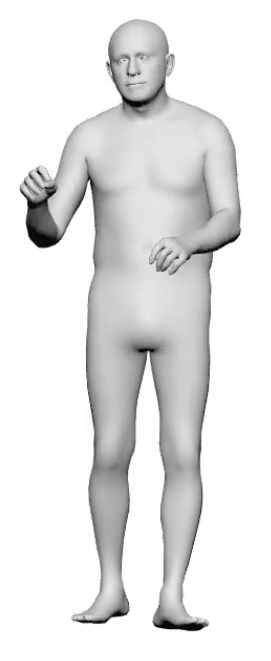}\\[-0.5mm]

        \methodcell{EMAGE} &
        \imgcell{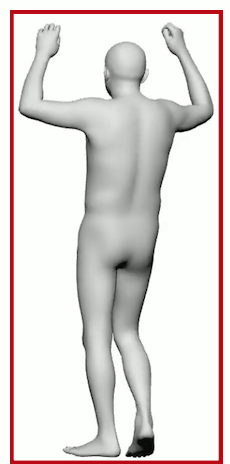} &
        \imgcell{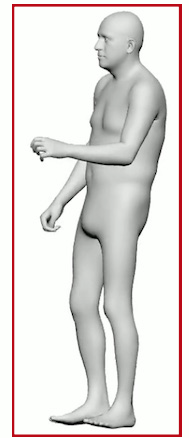} &
        \imgcell{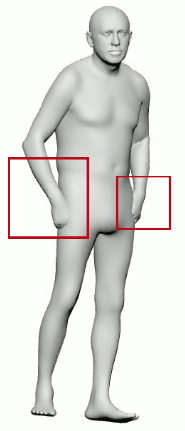} &
        \imgcell{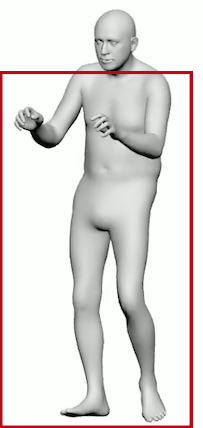} &
        \imgcell{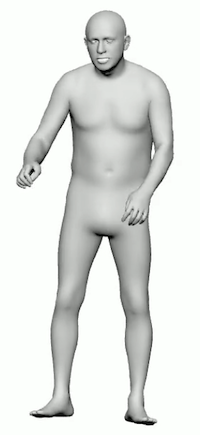}&
        \imgcell{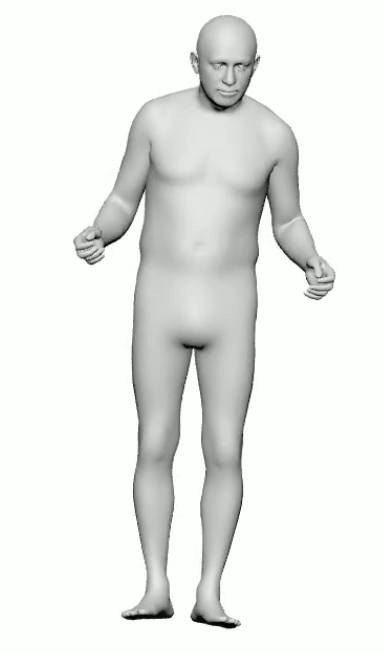}&
        \imgcell{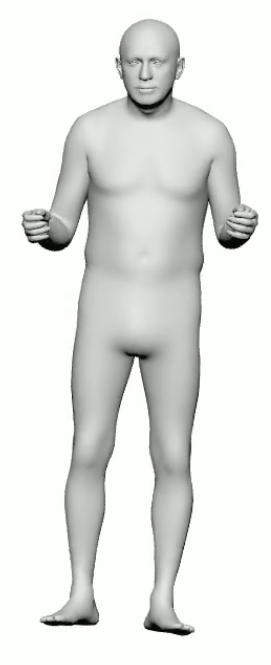}\\[-0.5mm]

        \methodcell{MambaTalk} &
        \imgcell{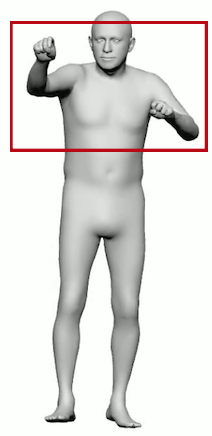} &
        \imgcell{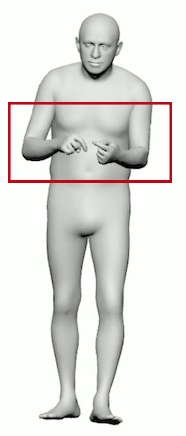} &
        \imgcell{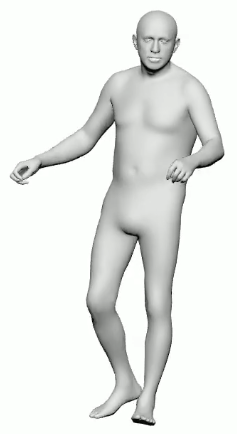} &
        \imgcell{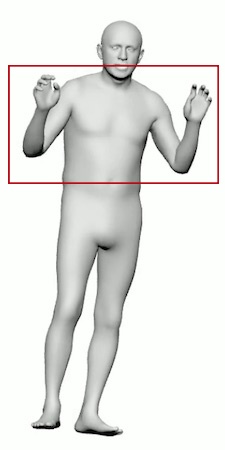} &
        \imgcell{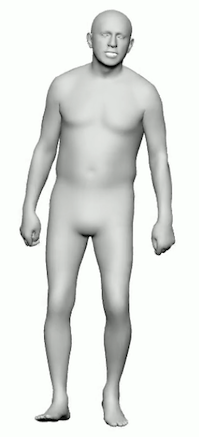}&
        \imgcell{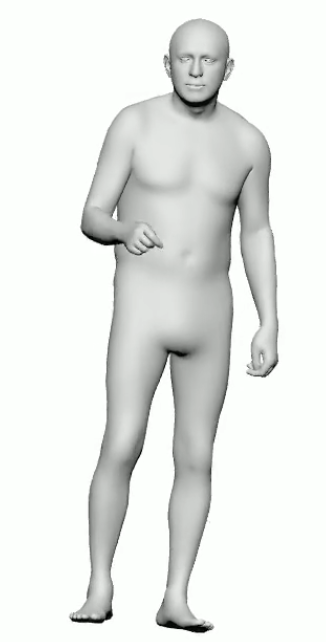}&
        \imgcell{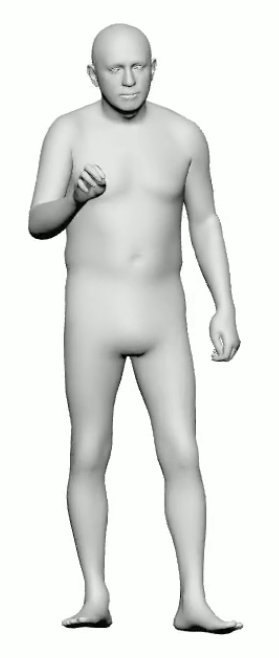}\\
    \end{tabular}%
    }

    \caption{ \textbf{Qualitative comparison on representative word-level speech contents.} The first row shows ground-truth motions and the corresponding speech contents, followed by results from different methods. Orange boxes highlight representative ground-truth motion regions, green boxes indicate generated motions closest to the corresponding references, and red boxes mark visible artifacts or semantically inconsistent motions. In the selected examples, EMAGE exhibits abrupt whole-body rotation on ``wow'' and ``especially,'' together with visible mesh deformation on ``in the street.'' These cases illustrate that larger-amplitude motions may still be accompanied by mesh deformation or temporal jitter. }
    \label{fig:qualitative_comparison_2}
\end{figure*}

We further compare body motion on 14 word-level cases in Fig.~\ref{fig:qualitative_comparison_1} and ~\ref{fig:qualitative_comparison_2}. Each case corresponds to the body pose generated at a specific spoken word, allowing direct comparison of gesture timing, body posture, and face--body coordination under the same speech context. Orange boxes highlight representative ground-truth motion regions used as visual references, green boxes indicate generated motions closest to the corresponding references, and red boxes mark visible artifacts or semantically inconsistent motions. In most selected cases, TalkFlow remains close to the ground truth while avoiding obvious artifacts such as implausible body deformation, abrupt large movements, or unstable poses.

\section{Conclusion}

We presented \textbf{TalkFlow}, a structured latent flow framework for speech-driven holistic motion generation. TalkFlow learns a continuous three-stream latent representation for upper-body motion, lower-body motion, and facial expressions, avoiding both direct high-dimensional motion generation and discrete tokenization. By encoding the full-body pose graph before temporal stream separation and coordinating the streams through lightweight FiLM-gated interactions, TalkFlow preserves part-specific dynamics while maintaining holistic face--body consistency.

We further introduced speech-conditioned latent flow matching to generate coordinated motion latents from acoustic and textual speech features. Experiments on BEAT2 show that TalkFlow achieves the best facial fidelity and competitive body-motion performance. Qualitative comparisons demonstrate plausible word-level lip dynamics and natural body gestures with fewer visible artifacts. The user study further shows that TalkFlow obtains the lowest mean rank in realism, semantic consistency, and speech-motion synchronization, together with the largest overall share of first-place rankings.

Several limitations remain. The single-speaker Speaker-2 protocol limits evidence for cross-speaker and cross-style generalization. The current Gaussian prior also lacks explicit talking-style control, limiting controllable diversity. Moreover, global translation is not modeled, making foot--ground contact difficult to constrain and contributing to occasional foot sliding. Future work will explore multi-speaker training, style-conditioned latent priors, and joint global-translation/contact constraints for more diverse and physically plausible motion.

\clearpage

\clearpage
\bibliographystyle{ACM-Reference-Format}
\bibliography{sample-base}

\end{document}